\documentclass[pre,preprint,english,showpacs,aps,prb,a4paper,floatfix]{revtex4}
\usepackage[T1]{fontenc}
\usepackage{color}
\usepackage{amsmath}
\usepackage[latin1]{inputenc}
\usepackage{graphicx}
\usepackage{epsfig}
\begin{document}

\title{Dimensional cross-over of hard parallel cylinders confined on cylindrical surfaces}

\author{Yuri Mart\'{\i}nez-Rat\'on}
\email{yuri@math.uc3m.es}
\affiliation{Grupo Interdisciplinar de Sistemas Complejos (GISC),
Departamento de Matem\'{a}ticas,Escuela Polit\'{e}cnica Superior,
Universidad Carlos III de Madrid, Avenida de la Universidad 30, E--28911, Legan\'{e}s, Madrid, Spain}

\author{Enrique Velasco}
\email{enrique.velasco@uam.es}
\affiliation{Departamento de F\'{\i}sica Te\'orica de la Materia Condensada
and Instituto de Ciencia de Materiales Nicol\'as Cabrera,
Universidad Aut\'onoma de Madrid, E-28049 Madrid, Spain}

\date{\today}

\begin{abstract}
We derive, from the dimensional cross-over criterion, a fundamental-measure density functional 
for parallel hard curved rectangles moving on a cylindrical surface.
We derive it from the density functional of circular arcs of length $\sigma$ with centers of mass 
located on an external circumference of radius $R_0$. 
The latter functional in turns is obtained from the corresponding 2D functional 
for a fluid of hard discs of radius $R$ on a flat surface with
centers of mass confined onto a circumference of radius 
$R_0$. Thus the curved length of closest approach between two centers of mass of hard discs on this circumference 
is $\sigma=2R_0\sin^{-1}(R/R_0)$, 
the length of the circular arcs.  
From the density functional of circular arcs, and by applying a dimensional expansion procedure to the spatial dimension 
orthogonal to the plane of the circumference, we finally obtain the density functional of curved rectangles of edge-lengths 
$\sigma$ and $L$. The DF for curved rectangles can also be obtained by fixing the centers of mass of parallel hard cylinders 
of radius $R$ and length $L$ on a cylindrical surface of radius $R_0$. 
Along the derivation, we show that when the centers of mass of the discs are confined to the exterior 
circumference of a circle of radius $R_0$: (i) for $R_0>R$, the exact Percus 1D-density functional of circular arcs of 
length $2R_0\sin^{-1}(R/R_0)$ is obtained, and (ii) for $R_0<R$, the 0D limit (a cavity that can hold one particle at most) 
is recovered.  
We also show that, for $R_0>R$, the obtained functional is equivalent to that 
of parallel hard rectangles on a flat surface of the same 
lengths, except that now the density profile of curved rectangles is a periodic function of the azimuthal angle, 
$\rho(\phi,z)=\rho(\phi+2\pi,z)$. The phase behavior of a fluid of aligned curved rectangles is obtained by calculating the 
free-energy branches of smectic, columnar and crystalline phases for different values of the ratio $R_0/R$ in the range 
$1<R_0/R\leq 4$; the smectic 
phase turns out to be the most stable except for $R_0/R=4$ where the crystalline phase becomes reentrant in a small range 
of packing fractions. 
When $R_0/R<1$ the transition is absent, since the 
density functional of curved rectangles reduces to the 1D Percus functional.  
\end{abstract}


\maketitle
\section{Introduction}

The subjects of surfaces decorated with particles in periodically or quasiperiodically packed configurations and the arrangement
of spheres on spherical, cylindrical or general surfaces have attracted a long-standing
interest \cite{Nelson0,Nelson}. The focus has been mainly on the type of packing, defect stabilisation and interactions, and the
topological constraints associated with non-vanishing Gaussian curvature. A most prominent problem concerns the stabilisation of 
crystalline order on a sphere, first predicted \cite{Bowick} and then experimentally observed in colloidal spheres on water droplets 
\cite{Bausch}. The problem has many important consequences in a number of fields; for example, possible packings of protein capsomeres 
in spherical viruses \cite{Lidmar}, behaviour of particles inserted in lipid membranes and their curvature-induced 
interactions \cite{Sachdev,Dinsmore}, metallic clusters, structure of fullerenes, to name just a few. The spherical or effectively spherical
surface is the most studied, while several studies have also appeared on cylindrical surfaces. Recently, Nelson studied the 
interaction of dislocations on a cylindrical surface \cite{Nelson2}; despite the vanishing Gaussian curvature of the cylinder, a rich
phenomenology was found.
  
Of particular interest is the case when particles are not spherical but elongated, since here there are issues of packing not only from 
the translational but also from the orientational degrees of freedom. Hence one has here two fields, the position and 
the nematic-director fields, that compete and interact with the geometry to possibly stabilise complex defected patterns. In a nematic phase 
particle positions are disordered but the nematic director is still contrained by the topology. The possibility of stabilising
defects by confining a thin nematic film on a spherical surface is intriguing.
For example, using computer simulation, Dzubiella et al. \cite{Dzubiella} studied the nematic ordering of hard rods confined to be on 
the surface of a sphere. In this case the confinement on a spherical surface induces a global topological charge in 
the director field due to the stabilisation of half-integer topological point defects.

Another interesting, much less analysed, aspect, concerns the formation of phases with partial positional order (liquid-crystalline)
from a disordered phase and the conditions imposed by topology, curvature and periodicity on the ordering and phase interplay as 
thermodynamic conditions (such as surface density of particles or temperature) are varied. In liquid-crystalline phases with partial or 
total translational order (smectic, columnar or crystalline) the two fields are coupled, and a complex interaction with the topology 
may result. This problem may be relevant in connection with the ordering of large protein molecules inserted into lipid membranes, where
curvature may both induce or modify order and be induced or modified by order. 

In the present article we focus on a system of parallel circular, rectangular particles confined to a cylindrical surface (therefore
the nematic phase is the most disordered phase of the system). This
is motivated as a useful model to discuss ordering of squared or rectangular proteins or otherwise on rod-shaped bacterial cell 
membranes, but the model can be analysed in a broader context. Despite its simple topology, the non-vanishing 
curvature and periodicity perpendicular to the cylinder axis may induce or supress ordering in some of the two orthogonal
directions, giving rise to possible smectic or columnar orderings of the particles on the cylindrical surface. This problem
has some similarities with the adsorption of particles {\it inside} slit-like or cylindrical pores (an example of which is
the recent study on the confinement of hard spheres (HS) into cylindrical pores \cite{White}, or of hard rods into planar pores
\cite{kike1,kike2}, both employing the density-functional (DF) formalism).  
Recently the close-packed structures of HS confined 
in cylindrical pores of small radii were classified using analytical methods and computer simulations \cite{Mughal}. 
All of these studies reflect the importance of the
commensuration between the pore width (in our case the circle diameter) and the characteristic dimensions of particles in the structure and 
stability of the confined non-uniform phases. Recent studies show that the extreme confinement of particles along one direction 
makes the system behave like an ideal gas in this direction. Thus the corresponding degrees of freedom can be integrated out 
and the resulting lower-dimensional system can be described by an effective interparticle potential \cite{Franosch}.
 
The purpose of the present article is twofold. First, we derive a density-functional theory for a fluid of
parallel circular rectangles on an external cylindrical surface; this model is isomorphic to a fluid monolayer of parallel cylinders 
adsorbed on an external cylindrical surface, or to the same fluid confined between two concentric cylindrical surfaces such that only 
one shell of cylinders can be accommodated with no radial motion. We show that the functional may be obtained consistently from different 
routes due to the important dimensional-crossover property of the theory. This property was first used to derive a fundamental measure 
density functional (FMF) for HS \cite{Schmidt,tarazona1} and parallel hard cubes (PHC) \cite{cuesta1}, 
and it was recently applied to obtain 
a density functional for hard parallel cylinders \cite{yuri2}. Second, the model is analysed statistical-mechanically
by investigating the free-energy minima landscape. We discuss different r\'egimes for the ratio of radius of the external cylinder 
to radii of the underlying adsorbed cylinders. When the ratio is sufficiently low the model is an effectively 1D model and no phase
transition exists. Otherwise the smectic phase is found to be the most stable except for a relative small range of densities in 
which the crystalline phase appears as a reentrant phase. These results are in line with those of recent experiments on confined 
liquid crystals in silica-glass nanochannels, which show that the stability of the smectic phase is considerably enhanced by the confinement 
at the expense of 
the crystalline phase \cite{Andri}. We finally propose, following the dimensional cross-over property,
a DF for spherical lenses (the intersection between a HS of radius 
$R$ which center of mass is located on an external sphere of radius $R_0$, with $R_0>R$) moving on a spherical surface. 

The article is arranged as follows. In Section \ref{SecI} we derive the density-functional theory for hard curved rectangles (CR)
on an external cylindrical surface. This is done in two steps: in the first (Section \ref{SecIa}), the 2D functional for hard discs
(HD) is projected on the surface of an external circle, thus obtaining a 1D functional for curved arcs (CA) that move along the circumference 
of the external circle. In the second (Section \ref{SecIb}) the functional is developed along the direction perpendicular to the
circle to give a 2D functional for CR. Dimensional consistency is discussed in Section \ref{SecIc}. The results are presented in
Section \ref{SecII}. In Section \ref{lenses} we propose, following the same dimensional cross-over recipe, a DF 
for spherical lenses which center of mass are confined on a external sphere. Finally  
some conclusions are presented in Section \ref{SecIII}. Details on the derivation of the density functionals 
and the density 
profile parameterizations are relegated to Appendices \ref{a_a}-\ref{a_e}.  

\section{Derivation of functionals}
\label{SecI}

In this section we derive the FMF for our model system, i.e. a fluid of CR. We do it in two steps. First,
the 2D functional for HD of radius $R$ is projected on the external surface of a cylinder of radius $R_0$, providing a 1D functional
for a fluid of CA (Section \ref{SecIa}). Then, this 1D functional is developed along the orthogonal dimension to obtain a 2D functional 
for the CR fluid (Section \ref{SecIb}). The two cases $R_0<R$ and $R_0>R$ are discussed separately in each case, since they give rise 
to fundamentally different expressions. In the last part of this section the resulting functionals are shown to be dimensionally 
consistent.

\subsection{Functional for CA}
\label{SecIa}
We start from the FMF excess free energy for a fluid of HD derived in Refs. \cite{tarazona1,yuri2}, i.e.
\begin{eqnarray}
\beta{\cal F}_{\rm ex}^{(\rm HD)}[\rho]=\int d{\bf r} \Phi^{(\rm HD)}_{\rm 2D}({\bf r}),
\label{excess}
\end{eqnarray}
where ${\bf r}=(r,\phi)$ is the radius vector in polar coordinates. The reduced free-energy density is defined as
\begin{eqnarray}
\Phi^{(\rm HD)}_{\rm 2D}({\bf r})=-n_0^{(\rm HD)}({\bf r})\ln\left[1-\eta_{\rm HD}({\bf r})\right]+
\frac{N_{\rm HD}({\bf r})}{1-\eta_{\rm HD}({\bf r})},
\label{oo}
\end{eqnarray}
where the weighted densities are convolutions of the two-dimensional density profile $\rho_{\rm 2D}({\bf r})$: 
\begin{eqnarray}
&&n_0^{(\rm HD)}({\bf r})=\frac{1}{2\pi R}\int d{\bf r}_1\rho_{\rm 2D}({\bf r}_1)\delta\left(R-|{\bf r}-{\bf r}_1|\right), 
\label{n0}\\\nonumber\\
&&\eta_{\rm HD}({\bf r})=\int d{\bf r}_1\rho_{\rm 2D}({\bf r}_1)\Theta\left(R-|{\bf r}-{\bf r}_1|\right),\label{local}\\\nonumber\\
&&N_{\rm HD}({\bf r})=\frac{1}{(2\pi R)^2}\int d{\bf r}_1\int d{\bf r}_2\rho_{\rm 2D}({\bf r}_1)\rho_{\rm 2D}({\bf r}_2) {\cal K}(r_{12})
\delta\left(R-|{\bf r}-{\bf r}_1|\right)\delta\left(R-|{\bf r}-{\bf r}_2|\right),\nonumber\\ 
\label{two-body}
\end{eqnarray}
with $\delta(x)$ and $\Theta(x)$ the Dirac-delta and Heaviside functions, respectively. Note that $\eta_{\rm HD}({\bf r})$ 
is just the local packing fraction, while $N_{\rm HD}({\bf r})$ is a two-body weighted density defined through the kernel
\begin{eqnarray}
{\cal K}(r)=4\pi R^2 x\sin^{-1}(x)\sqrt{1-x^2}\Theta(1-x),\quad x\equiv\frac{r}{2R}.
\label{kernel}
\end{eqnarray}

\begin{figure}
\epsfig{file=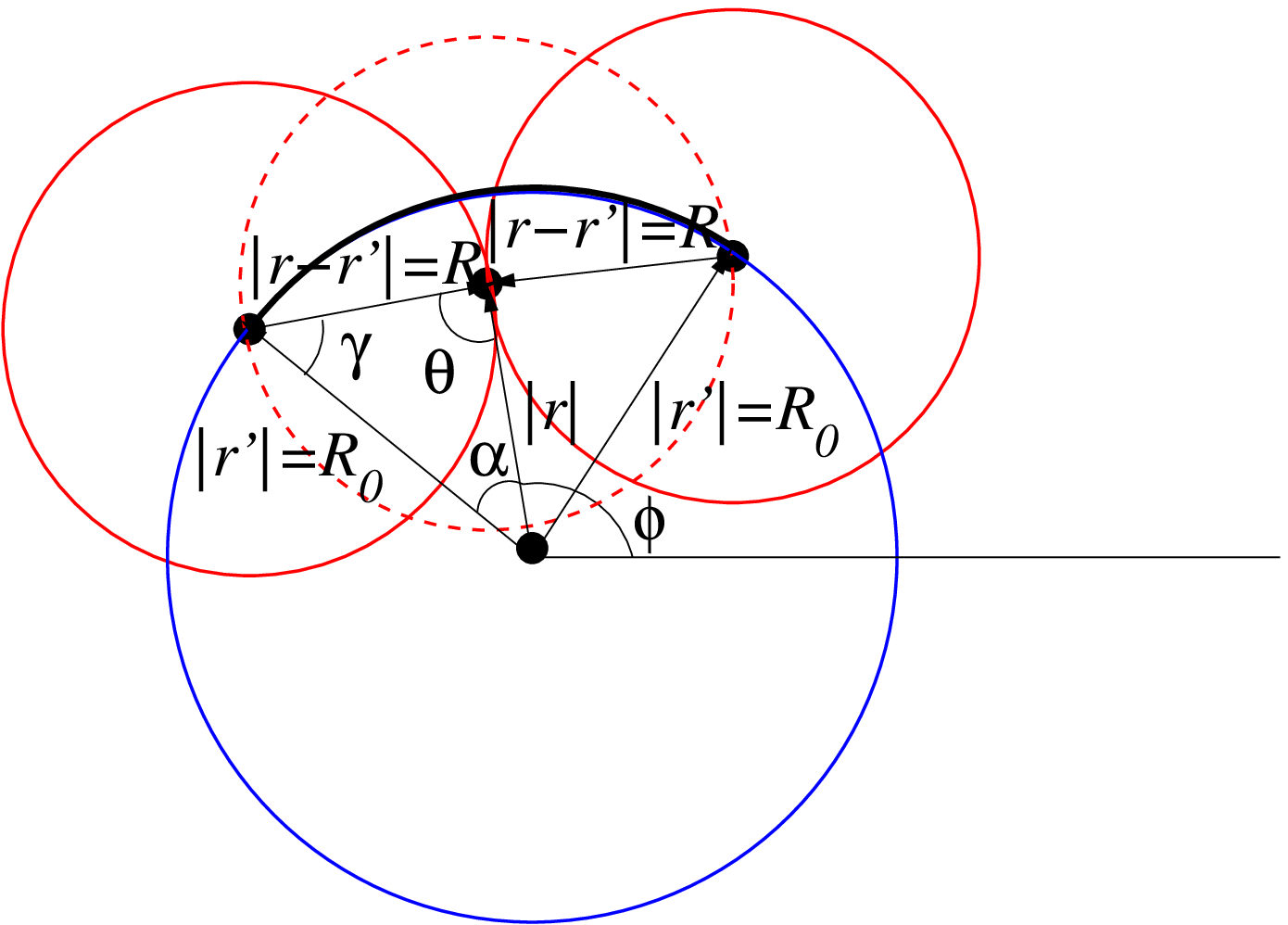,width=3.in}
\epsfig{file=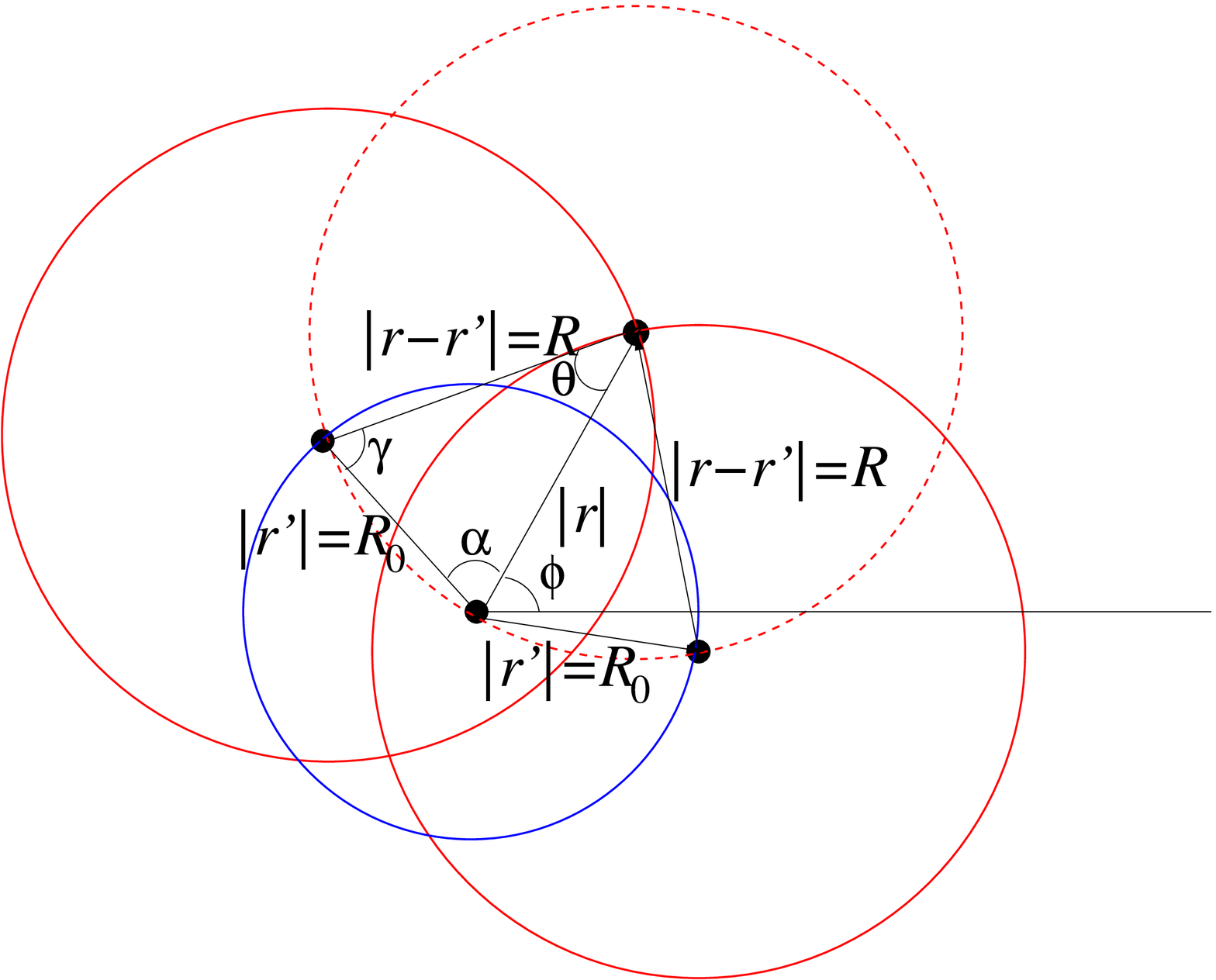,width=3.in}
\caption{(Color online) Sketch of the HDs configuration on a circle with $R_0>R$ (left) and $R_0<R$ (right).}
\label{fig1}
\end{figure}

We now restrict the degrees of freedom by imposing that the HD centers of mass be located on a circumference of radius $R_0$:
\begin{eqnarray}
\rho_{\rm 2D}(r,\phi)=\rho_{\rm 1D}(\phi)\delta(R_0-r),
\label{rho}
\end{eqnarray}
 where 
$\rho_{\rm 1D}(\phi)$ is the 1D density profile, and substitute Eq. (\ref{rho}) into Eqs. (\ref{n0}), (\ref{local}) 
and (\ref{two-body}). Then we take the following steps (details of which can be found in Appendix \ref{a_a}):
(i) A first change of variables $(\phi_i,r_i)\to(\phi_i,\xi_i)$, $i=1,2$, with $\xi_i=|{\bf r}-R_0{\bf u}_i|$ 
[${\bf u}_i=(\cos\phi_i,\sin\phi_i)$ are unit vectors] to evaluate the weighted densities. As a result they become 
functions solely of $\phi$ and $r$. (ii) Express the weighted densities as a function of the three angles 
$\alpha(r)$, $\theta(r)$ and $\gamma(r)$ [the inner angles of the triangle of sides $r$, $R$ and $R_0$, see Fig. \ref{fig1}]
and of their derivatives with respect to $r$. (iii) Write the second term of the expression for the free-energy density 
(\ref{oo}) as a sum of the derivatives of the first term with respect to $\alpha$ and $\phi$. (iv) A second change of variables 
$(\phi,r)\to(\phi,\gamma)$ in Eq. (\ref{excess}). (v) Use of the periodicity of 
the density profile with respect to the azimuthal angle, $\rho(r,\phi)=\rho(r,\phi+2\pi)$, and integration by parts 
to finally arrive at the following expression for the excess part of the DF of HD: 
\begin{eqnarray}
\beta{\cal F}_{\rm ex}^{(\rm HD)}[\rho]=\frac{R_0}{\pi}\int_0^{2\pi}d\phi\int_0^{\pi}d\gamma\left\{
\Psi(\phi,\alpha(\gamma))+\frac{\partial \Psi}{\partial\alpha}(\phi,\alpha(\gamma))
\theta^*(\gamma)\left|\frac{d\alpha}{d\gamma}(\gamma)\right|\right\}.
\label{ener}
\end{eqnarray}
Here the shorthand notations 
$\Psi(\phi,\alpha)=-n_{0+}(\phi,\alpha)\ln\left[1-\eta_{\rm HD}(\phi,\alpha)\right]$ and 
$n_{0+}(\phi,\alpha)=[\rho_{\rm 1D}(\phi+\alpha)+\rho_{\rm 1D}(\phi-\alpha)]/2$ were used, while the local packing 
fraction and the angle $\theta^*$ are defined as 
$\eta_{\rm HD}(\phi,\alpha)=R_0\int_{\phi-\alpha}^{\phi+\alpha}d\phi_1\rho_{\rm 1D}(\phi_1)$ and 
$\theta^*=\theta$ if $0\leq\theta\leq \pi/2$, while $\theta^*=\pi/2-\theta$ if $\pi/2<\theta\leq\pi$.

\subsubsection{The case $R_0>R$}

A look at Fig. \ref{fig1} allows us to write the following two relations between the angles $\alpha$ and $\gamma$:
\begin{eqnarray}
\displaystyle{\frac{d\alpha}{d\gamma}\geq 0}\hspace{0.4cm}\text{if}\hspace{0.4cm} 0<\gamma\leq \gamma_0,\hspace{0.6cm}
\displaystyle{\frac{d\alpha}{d\gamma}<0}\hspace{0.4cm}\text{if}\hspace{0.4cm} \gamma_0<\gamma\leq \pi,
\label{angulos}
\end{eqnarray}
where $\gamma_0=\cos^{-1}(R/R_0)$, $\theta_0\equiv\theta(\gamma_0)=\pi/2$ and $\alpha_0\equiv\alpha(\gamma_0)=\sin^{-1}(R/R_0)$.
After some lengthy calculations, which can be followed in detail in Appendix \ref{a_b}, we obtain from (\ref{ener}) and (\ref{angulos}) 
the following dimensional cross-over:
\begin{eqnarray}
 \beta{\cal F}^{(\rm HD)}_{\rm ex}[\rho]\to \beta{\cal F}^{(\rm CA)}_{\rm ex}[\rho]=R_0\int_0^{2\pi}d\phi\Phi_{\rm 1D}^{(\rm CA)}(\phi),
 \label{limite1}
\end{eqnarray}
with $\Phi^{\rm CA}_{\rm 1D}(\phi)$ the 1D Percus free-energy density:
\begin{eqnarray}
\Phi^{(\rm CA)}_{\rm 1D}(\phi)\equiv\Psi(\phi,\alpha_0)=-n_0^{(\rm CA)}(\phi)\ln\left[1-\eta_{\rm CA}(\phi)\right],
\label{CA}
\end{eqnarray}
and with the corresponding weighted densities
\begin{eqnarray}
&&n_0^{(\rm CA)}(\phi)=n_{0+}(\phi,\alpha_0)=\frac{1}{2}\left[\rho_{\rm 1D}(\phi-\alpha_0)+\rho_{\rm 1D}(\phi+\alpha_0)\right],
\label{redefine_1}\\\nonumber\\
&&\eta_{\rm CA}(\phi)=\eta_{\rm HD}(\phi,\alpha_0)=R_0\int_{\phi-\alpha_0}^{\phi+\alpha_0}d\phi'\rho_{\rm 1D}(\phi').
\label{redefine_2}
\end{eqnarray}
Here the index `CA' means that these densities are evaluated on a circular arc of length $2R_0\alpha_0$. Then we have proved 
that the excess part of the HD free-energy functional reduces to that of hard CA.

\subsubsection{The case $R_0<R$}

In this case we always have 
\begin{eqnarray}
\frac{d\alpha}{d\gamma}<0,\quad \forall \gamma,\quad 0\leq \theta <\pi/2
\label{angles2}
\end{eqnarray}
(see Fig. \ref{fig1}). After some algebra (described in detail in Appendix \ref{a_c}), we obtain from (\ref{ener}) and (\ref{angles2}) 
the following  dimensional cross-over:
\begin{eqnarray}
\beta {\cal F}_{\rm ex}^{(\rm HD)}[\rho]\to \beta {\cal F}_{\rm ex}^{(\rm CA)}[\rho]=\eta_{\rm CA}+(1-\eta_{\rm CA})\ln(1-\eta_{\rm CA}),
\label{0DD}
\end{eqnarray} 
corresponding to the exact 0D functional of CA (such that at most one arc can exist on the circle).
The mean number of particles is
\begin{eqnarray}
\eta_{\rm CA}\equiv 
R_0\int_{-\pi}^{\pi}d\phi'\rho_{\rm 1D}(\phi')=2\pi R_0 \overline{\rho}_{\rm 1D},
\end{eqnarray}
with $\overline{\rho}_{\rm 1D}$ the mean number density over the circle of radius $R_0$.

\subsection{Functional for CR}
\label{SecIb}

We now define a collection of CR, each consisting of two parallel curved edges formed by circular arcs of 
length $2R_0\alpha_0$ and two parallel straight lines of length $L$ perpendicular to the former. The centers of mass 
of the CR are located on a 2D cylindrical surface of radius $R_0$. The density profile will be $\rho_{\rm 2D}(\phi,z)$, 
where $z$ is the coordinate along the cylinder axis and $\phi$ is the azimuthal angle. The functional for this model
will be obtained in the two cases. 

\subsubsection{The case $R_0>R$}

First, we define the local packing fraction as
\begin{eqnarray}
\eta_{\rm CR}(\phi,z)=R_0\int_{\phi-\alpha_0}^{\phi+\alpha_0}d\phi'\int_{z-L/2}^{z+L/2}dz'\rho_{\rm 2D}(\phi',z'),
\label{pack}
\end{eqnarray}
and the weighted density
\begin{eqnarray}
n_{1,\perp}^{(\rm CR)}(\phi,z)=\frac{1}{2}\int_{z-L/2}^{z+L/2}dz'\left[\rho_{\rm 2D}(\phi-\alpha_0,z')+
\rho_{\rm 2D}(\phi+\alpha_0,z')\right].
\label{modified}
\end{eqnarray}
Let us write the modified 1D Percus free-energy density:
\begin{eqnarray}
\tilde{\Phi}_{\rm 1D}(\phi,z)=-n_{1,\perp}^{(\rm CR)}(\phi,z)\ln\left[1-\eta_{\rm CR}(\phi,z)\right].
\end{eqnarray}
Note that this is in fact a local free-energy density corresponding to the CA fluid.
Now the CR free-energy density can be calculated by applying a differential operator $\partial/\partial L$ to 
the modified 1D free-energy density \cite{cuesta1}:
\begin{eqnarray}
\Phi_{\rm 2D}^{(\rm CR)}(\phi,z)=\frac{\partial}{\partial L}\tilde{\Phi}_{\rm 1D}(\phi,z),
\label{cr}
\end{eqnarray}
which results in
\begin{eqnarray}
\Phi_{\rm 2D}^{(\rm CR)}(\phi,z)=-n_0^{(\rm{CR})}(\phi,z)\ln\left[1-\eta_{\rm CR}(\phi,z)\right]+
\frac{n_{1,\perp}^{(\rm CR)}(\phi,z)n_{1,\parallel}^{(\rm CR)}(\phi,z)}{1-\eta_{\rm CR}(\phi,z)},
\label{phi_cr}
\end{eqnarray}
where we have defined
\begin{eqnarray}
&&\hspace{-0.4cm}n_0^{(\rm CR)}(\phi,z)=\frac{\partial}{\partial L} \tilde{n}_{1,\perp}^{(\rm CR)}(\phi,z)=
\frac{1}{4}\left[\rho_{\rm 2D}(\phi_-,z_-)+\rho_{\rm 2D}(\phi_-,z_+)
+\rho_{\rm 2D}(\phi_+,z_-)+\rho_{\rm 2D}(\phi_+,z_+)\right],\nonumber\\\label{nn0}\\
&&\hspace{-0.4cm}n_{1,\parallel}^{(\rm CR)}(\phi,z)=\frac{\partial}{\partial L}\eta_{\rm CR}(\phi,z)=
\frac{R_0}{2}\int_{\phi_-}^{\phi_+}d\phi'\left[\rho_{\rm 2D}(\phi',z_-)+
\rho_{\rm 2D}(\phi',z_+)\right]\label{nnpar},
\end{eqnarray}
and $z_{\pm}=z\pm L/2$, $\phi_{\pm}=\phi\pm\alpha_0$.
Then the excess CR free-energy functional can be calculated as
\begin{eqnarray}
\beta {\cal F}_{\rm ex}^{(\rm CR)}[\rho]=R_0\int_0^{2\pi}d\phi\int_{-\infty}^{\infty}dz\Phi_{\rm 2D}^{(\rm CR)}(\phi,z).
\label{finalizo}
\end{eqnarray}
Note that (\ref{phi_cr}) is just the excess free-energy density of a fluid of parallel 
hard rectangles (PHR) of length $L$ and width $\sigma=2R_0\alpha_0$ which in turn coincides with that of the 
parallel hard squares (PHS) fluid after scaling one of the edge-lengths \cite{cuesta1}. Also note that, taking the 
limit $R_0/R\to\infty$, changing the variable $\phi$ to $x\equiv R_0\phi$ and setting $\sigma=2R_0\alpha_0$, we obtain from (\ref{finalizo}) the excess part of the density functional of PHR on a flat surface 
of edge-lengths $\sigma$ and $L$  given in Ref. \cite{cuesta1}. Thus the present functional 
has the same degree of exactness as that for the PHS fluid. As was shown in \cite{Rene}, the minimization of the latter 
gives a phase behavior in which the columnar and crystalline phases both bifurcates from the uniform fluid branch at $\eta\approx 0.54$, 
with the columnar phase being the stable phase (although the difference between free energies is very small) up 
to $\eta\approx 0.73$, where a weak first-order columnar-to-crystal transition occurs. 
Single-speed Molecular Dynamics simulations of PHS show a second order melting transition at $\eta\approx 0.79$ \cite{Hoover}
similar to the above given value. However, columnar ordering was not found in the simulations.    

\subsubsection{The case $R_0<R$}

If $R_0<R$ we define a 1D local packing fraction as 
\begin{eqnarray}
\eta_{\rm CR}(z)=R_0\int_{-\pi}^{\pi}d\phi\int_{z-L/2}^{z+L/2}dz'\rho_{\rm 2D}(\phi',z'),
\end{eqnarray}
and a modified 0D free-energy density as 
\begin{eqnarray}
\tilde{\Phi}_{\rm 0D}^{(\rm CR)}(z)=\eta_{\rm CR}(z)+\left[1-\eta_{\rm CR}(z)\right]
\ln\left[1-\eta_{\rm CR}(z)\right].
\end{eqnarray}
The CR free-energy density for this case can be obtained by using the differential operator 
$\partial/\partial L$ applied to the modified 0D free-energy density  \cite{cuesta1}: 
\begin{eqnarray}
\frac{\partial}{\partial L}\tilde{\Phi}^{(\rm CR)}_{\rm 0D}(z)&=&
R_0\Phi_{\rm 1D}^{(\rm CR)}(z)=
-R_0n_0^{(\rm CR)}(z)\ln\left[1-\eta_{\rm CR}(z)\right],
\label{uy}
\end{eqnarray}
where 
\begin{eqnarray}
n_0^{(\rm CR)}(z)=R_0^{-1}\frac{\partial}{\partial L}\eta_{\rm CR}(z)=\frac{1}{2}\int_{-\pi}^{\pi}d\phi'
\left[\rho_{\rm 2D}(\phi',z_-)+\rho_{\rm 2D}(\phi',z_+)\right].
\end{eqnarray}
The excess free-energy functional is now
\begin{eqnarray}
\beta{\cal F}_{\rm ex}^{(\rm CR)}[\rho]=2\pi R_0\int_{-\infty}^{\infty}dz\Phi_{\rm 1D}^{(\rm CR)}(z).
\end{eqnarray}

\subsection{Confined hard cylinders}
\label{SecIc}

\begin{figure}
\epsfig{file=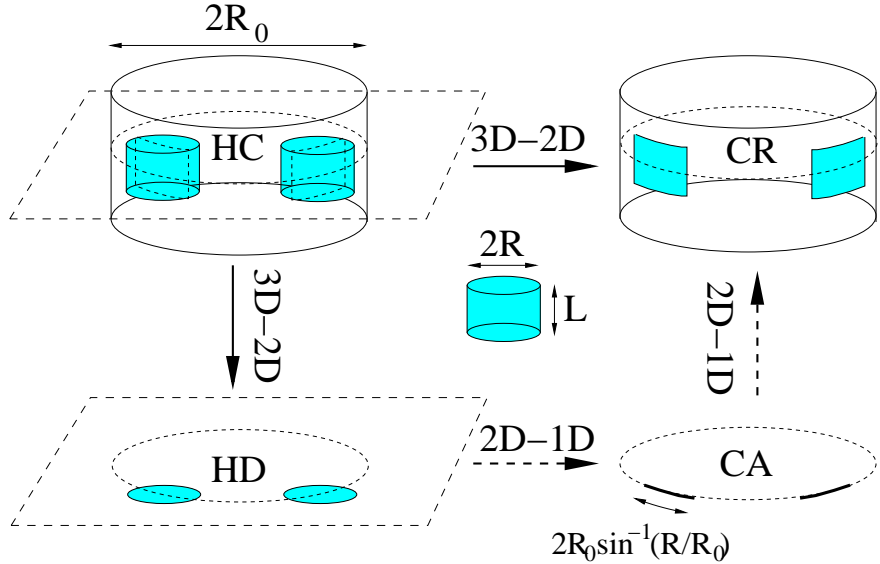,width=4.5in}
\caption{(Color online) Dimensional cross-over of HC of radius $R$ and length $L$ 
confined on a cylindrical surface of radius $R_0$. This $3D\to 2D$ crossover becomes HC into 
CR. Also shown is the $3D\to 2D$ cross-over when the HC are confined on a flat surface. 
This cross-over becomes HC into HD. Finally the $2D\to 1D$ cross-over from HD or CR when both 
are confined on a circumference of radius $R_0$ results in CA of length $2R_0\sin^{-1}(R/R_0)$ (the arc 
of closest approach between two HD). Dashed arrows indicate the route to obtain the FMF for CR.}
\label{cross-over}
\end{figure} 

The free energy of a collection of CR can be obtained from two different routes. 
In the previous two sections we used a two-step process: first, the HD fluid was projected on the circumference of a circle, 
giving the functional for CA. This in turn was developed along $z$ to give the free energy for CR. 
There is another possible route which starts from the 3D functional for hard cylinders (HC). The HC excess free-energy is
\begin{eqnarray}
\beta {\cal F}_{\rm ex}^{(\rm HC)}[\rho]=\int dz\int d\phi\int dr r\Phi_{\rm 3D}^{(\rm HC)}(r,\phi,z),
\label{ww}
\end{eqnarray}
with $\Phi_{\rm 3D}^{(\rm HC)}(r,\phi,z)$ the excess free-energy density \cite{yuri2,capi}. Now we confine the HC
on a plane perpendicular to the cylinder axes at $z=0$ by choosing the 3D density profile
$\rho_{\rm 3d}(r,\phi,z)=\rho_{\rm 2d}(r,\phi)\delta(z)$. Inserting this into (\ref{ww}),
we obtain an excess free-energy 
\begin{eqnarray}
\beta {\cal F}_{\rm ex}^{(\rm HD)}[\rho]=\int d\phi\int dr r\Phi_{\rm 2D}^{(\rm HD)}(r,\phi),
\end{eqnarray}
where, due to the dimensional crossover consistency, $\Phi_{\rm 2D}^{(\rm HD)}(r,\phi)$ is given by Eq. (\ref{oo}), i.e.
coincides with the HD free-energy density.
Selecting now $\rho(r,\phi,z)=\rho(\phi,z)\delta(R_0-r)$ (i.e. confining the HC on a 2D
cylindrical surface of radius $R_0$) and inserting it into Eq. (\ref{ww}), we obtain, for $R_0>R$ 
\begin{eqnarray}
\beta {\cal F}_{\rm ex}^{(\rm CR)}[\rho]=R_0\int_0^{2\pi}d\phi \int_{-\infty}^{\infty} dz \Phi_{\rm 2D}^{(\rm CR)}(\phi,z),
\end{eqnarray}
where $\Phi_{\rm 2D}^{(\rm CR)}(\phi,z)$ is the free-energy density of curved rectangles (\ref{cr}). When
$R_0<R$ we obtain
\begin{eqnarray}
\beta {\cal F}_{\rm ex}^{(\rm CR)}[\rho]=2\pi R_0\int_{-\infty}^{\infty} dz \Phi_{\rm 1D}^{(\rm CR)}(z),
\end{eqnarray}
where $\Phi_{\rm 1D}^{(\rm CR)}(z)$ is the free-energy density (\ref{uy}).

We sketch in Fig. \ref{cross-over} the dimensional cross-overs involved in the preceding discussion.

\section{Phase behaviour of CR}
\label{SecII}

In this section we analyze the phase behaviour of the CR fluid predicted by the functional derived in the previous section,
in the non-trivial case $R_0>R$. We start by presenting the numerical treatment in Section \ref{secA} and then show the results 
in Section \ref{secB}.
   
\subsection{Minimization technique}
\label{secA}

\begin{figure}
\epsfig{file=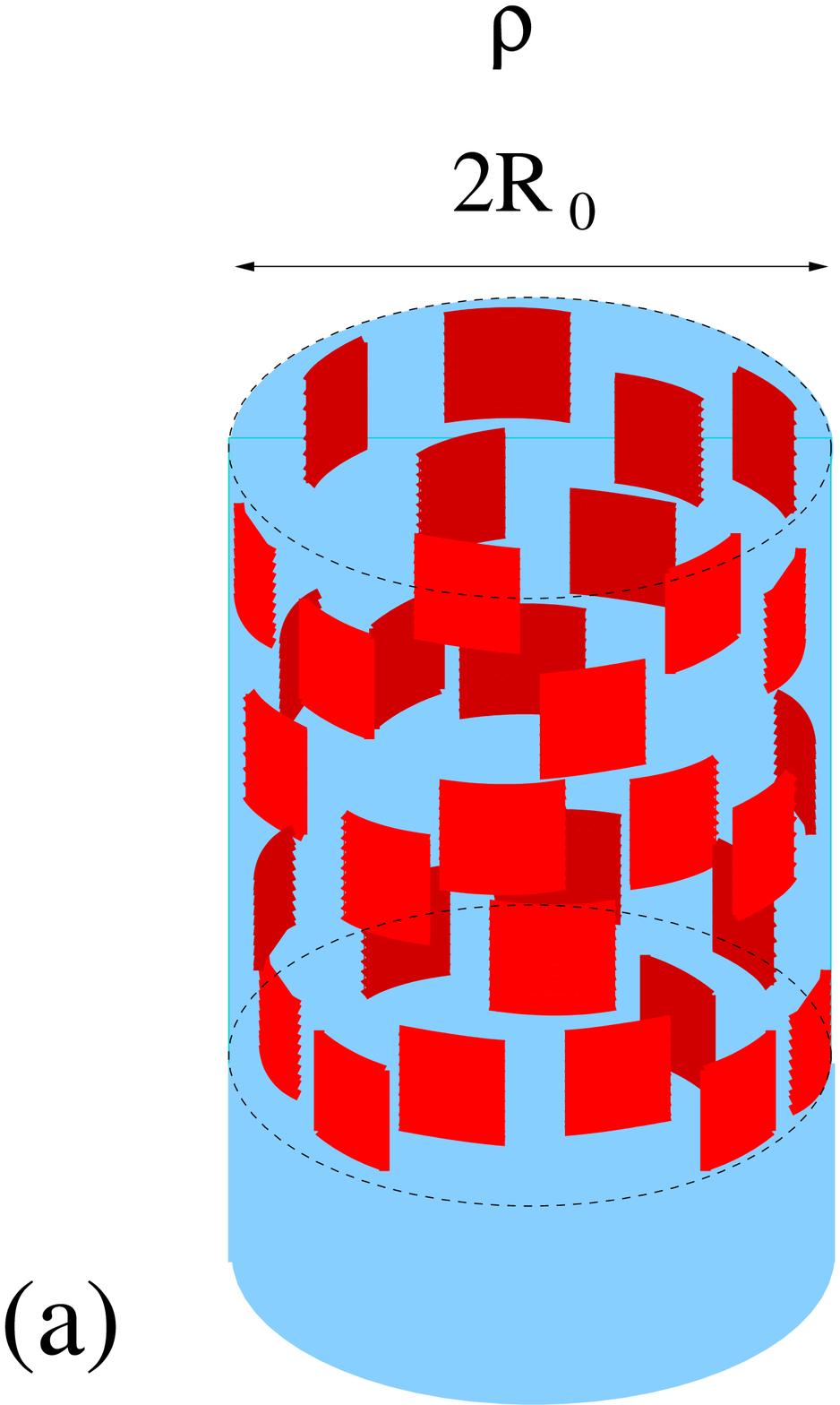,width=1.2in}
\epsfig{file=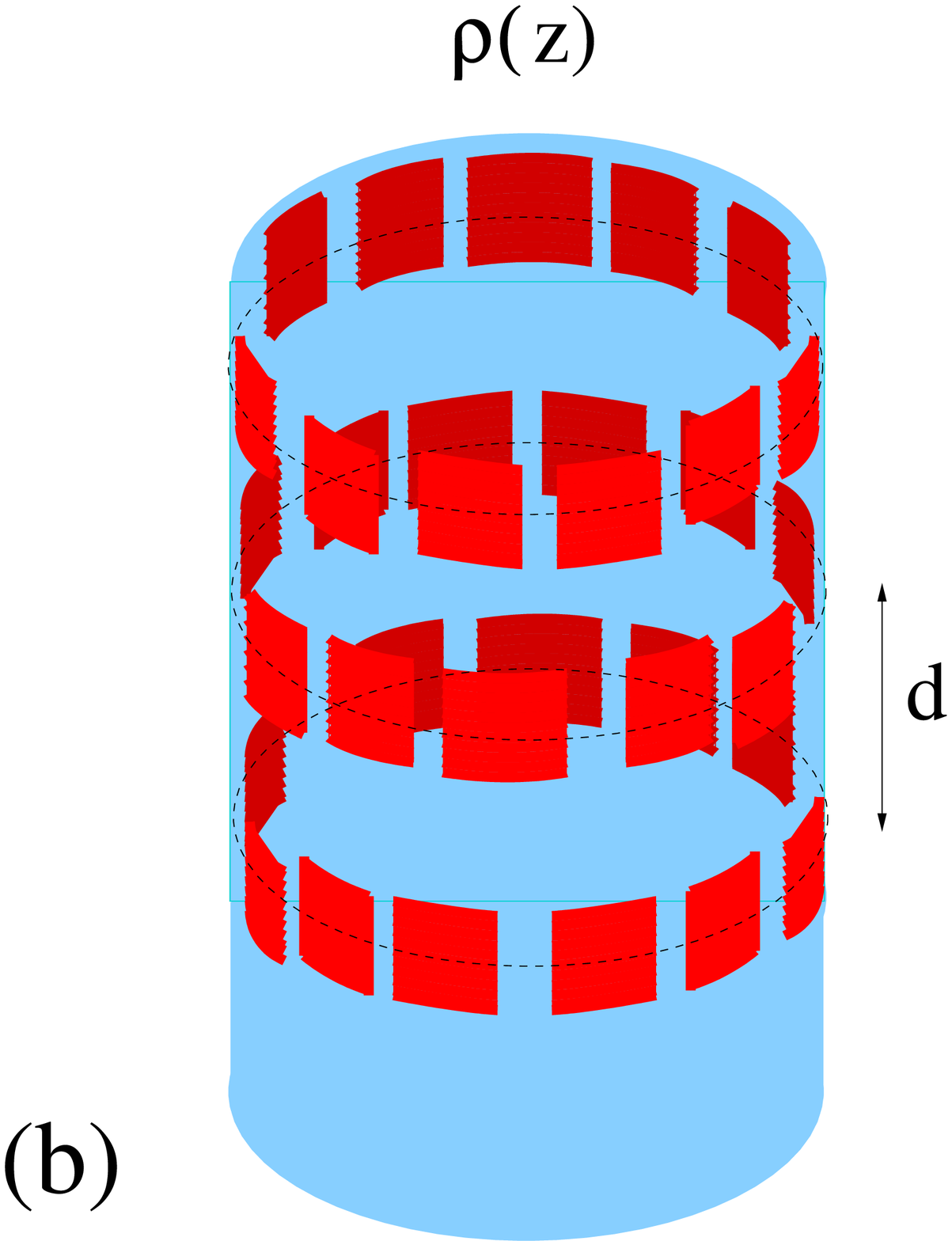,width=1.4in}
\epsfig{file=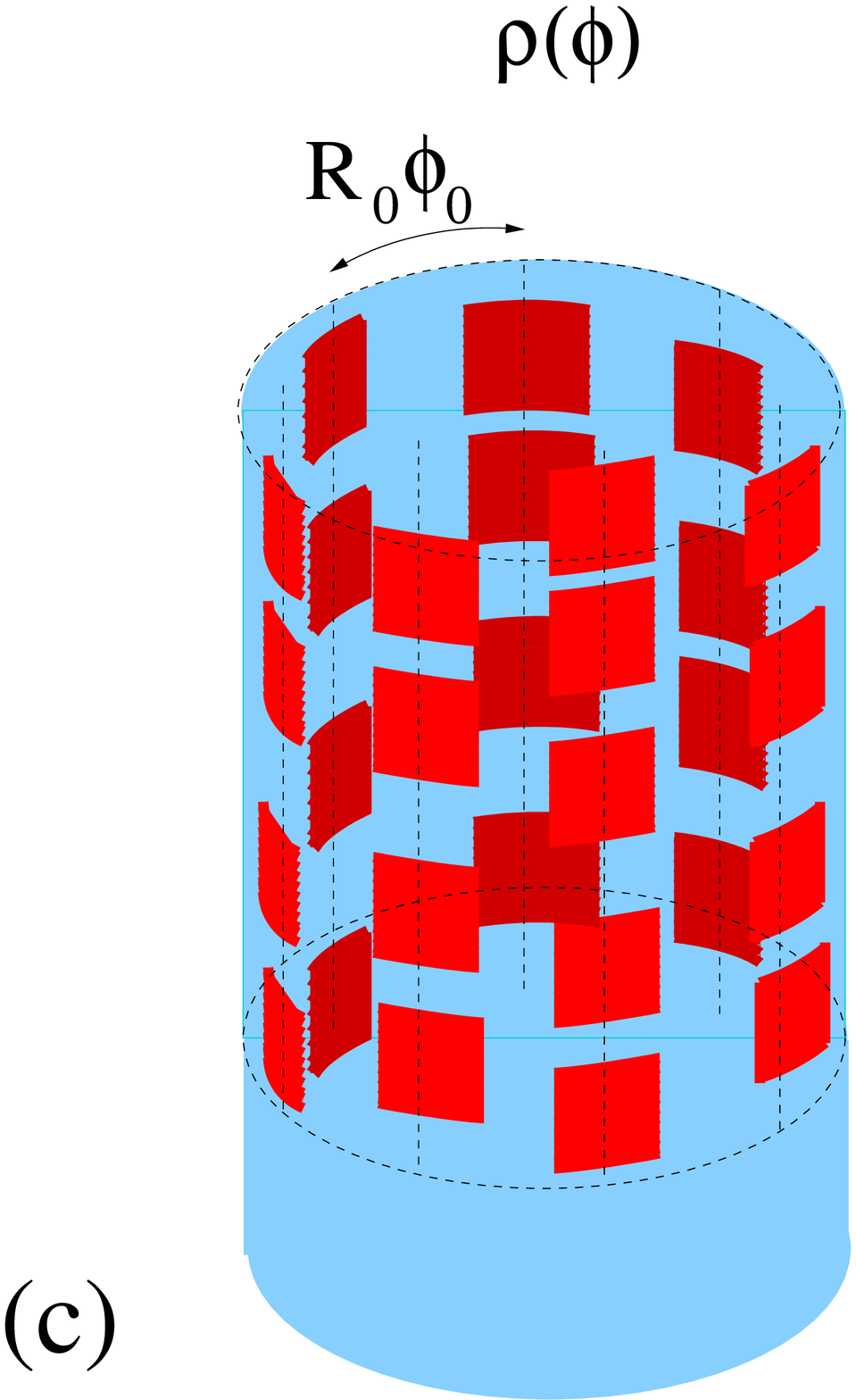,width=1.25in}
\epsfig{file=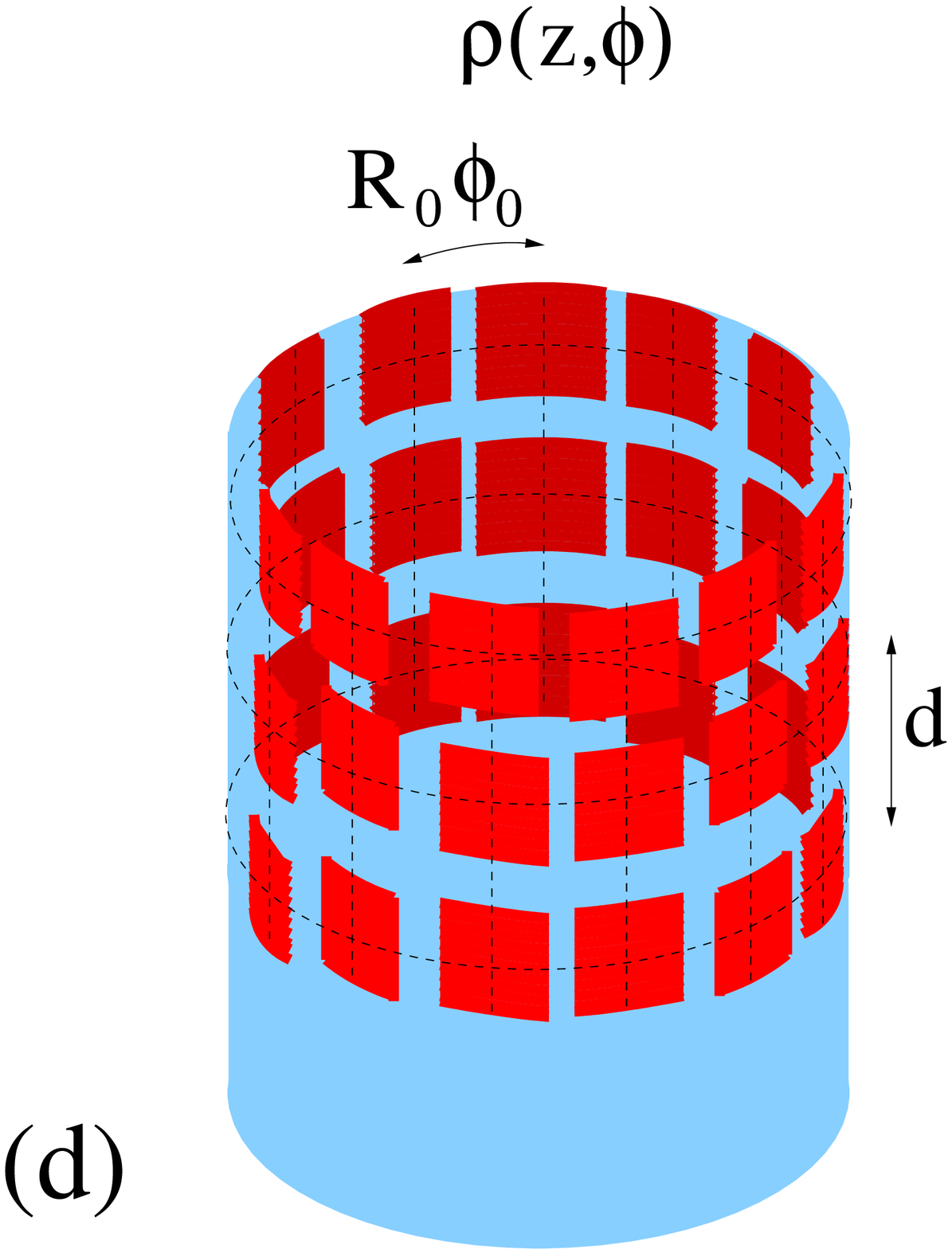,width=1.3in}
\caption{(Color online) Phases of CR on a cylindrical surface of radius $R_0$: 
(a) N, (b) S, (c) C and (d) K. The lattice parameters 
and the variables of which the density profiles depend are correspondingly labelled.}
\label{prueba}
\end{figure}

The total free-energy density functional (ideal plus excess parts) per unit of area is
\begin{eqnarray}
\frac{\beta {\cal F}^{(\rm CR)}[\rho]}{2\pi R_0 {\cal L}}=\frac{1}{2\pi d} \int_0^{2\pi}d\phi\int_0^d dz 
\left[\Phi_{\rm id}^{(\rm CR)}(\phi,z)+\Phi_{\rm 2D}^{(\rm CR)}(\phi,z)\right],
\label{todito}
\end{eqnarray}
where ${\cal L}$ is the system length along $z$ and
\begin{eqnarray}
\Phi_{\rm id}^{(\rm CR)}(\phi,z)=\rho_{\rm 2D}(\phi,z)\left[\log \rho_{\rm 2D}(\phi,z)-1\right],
\end{eqnarray}
is the ideal part. The excess contribution is given by Eq. (\ref{phi_cr}). The possible ordered phases in the system are
smectic (S), columnar (C) and crystal (K) (sketched in Fig. \ref{prueba}). 
A general density profile will be periodic in $z$ (with period $d$) and $\phi$
(with period $2\pi$), i.e. $\rho_{\rm 2d}(\phi+2\pi,z+d)=\rho_{\rm 2d}(\phi,z)$, and a convenient representation is a 
double Fourier expansion
\begin{eqnarray}
\rho_{\rm 2d}(\phi,z)=\rho_0\left[1+\sum_{(k,m)\neq (0,0)}s_{km} \cos ( kN_0\phi)\cos(q m z)\right],
\label{fourier}
\end{eqnarray}
where $\rho_0$ is the mean density $\rho_0=(2\pi d)^{-1}\int_0^{2\pi} d\phi \int_0^d dz \rho_{\rm 2d}(z,\phi)$, $\{s_{km}\}$ are the 
Fourier amplitudes, and $q=2\pi/d$ is the wave-vector along $z$. Note that $s_{km}=0$ $\forall$ $k\neq 0$ corresponds to the S phase,
while $s_{km}=0$ $\forall$ $m\neq 0$ implies a C phase. In the latter phase the obvious periodicity 
$\rho_{\rm 2d}(\phi)=\rho(\phi+2\pi)$ must be supplemented with a periodicity $\rho_{\rm 2d}(\phi+\phi_0)=\rho_{\rm 2d}(\phi)$, 
with $\phi_0=2\pi/N_0$ the period; here the average position of the columns would be located on the vertices of a $N_0$-sided 
regular polygon. $N_0$ is an integer in the interval 2--$[\pi/\alpha_0]$, with $[x]$ the integer part of $x$, and
$\alpha_0=\sin^{-1}(R/R_0)$. As an example, if $R_0=2R$ we have $N_0^{(\rm max)}=6$, 
a value that can be reached only at close-packing. 

Using the above Fourier expansion, the weighted densities (\ref{pack}), (\ref{modified})  and 
(\ref{nn0}), (\ref{nnpar})  can be calculated from the
expressions given in Appendix \ref{a_d} [see Eqs. (\ref{n0_f}-\ref{n1_par})]. The strategy followed to minimise the functional 
(\ref{todito}) was to use a truncated Fourier expansion in terms of the
amplitudes $\{s_{km}\}$ for $0\leq k,m\leq M$ (with $M$ selected in such a way as
to guarantee an adequate description of the density profile) and then minimise with respect to these amplitudes 
and the period $d$. We used a conjugate-gradient method to numerically implement the minimization. All the integrals were 
calculated using Legendre quadratures with as many roots as necessary to guarantee numerical errors in the amplitudes
below $10^{-7}$. 

As shown in the next section, the unstable or metastable character of the crystalline phase is difficult to obtain by direct
minimisation, so in this case we chose to parameterise the density profile as a sum of Gaussians: 
\begin{eqnarray}
&&\rho_{\rm 2D}(\phi,z)=(1-\nu)\zeta_{0\perp}(\phi)\zeta_{0\parallel}(z),\label{gaussian}\nonumber\\\nonumber\\
&&\zeta_{0\perp}(\phi)=\left(\frac{\Lambda_{\perp}}{\pi}\right)^{1/2}\sum_k\exp\left[-\Lambda_{\perp} R_0^2(\phi-k\phi_0)^2\right],
\nonumber\\\nonumber\\
&&\zeta_{0\parallel}(z)=\left(\frac{\Lambda_{\parallel}}{\pi}\right)^{1/2}\sum_k\exp\left[-\Lambda_{\parallel} (z-k d)^2\right],
\end{eqnarray}
where $\Lambda_{\perp} (2R_0\alpha_0)^2=\Lambda_{\parallel} L^2=\Lambda$ is the scaled Gaussian parameter and $\nu$ the fraction 
of vacancies.
Within this approximation, the total free energy per unit of area, in reduced units, can be computed from the expression 
given in Appendix \ref{a_d} [see Eq. (\ref{ooz})].  We minimized (\ref{ooz}) with respect to $\Lambda$ and $\nu$. The mean 
packing fraction can be computed as $\displaystyle{\eta_0=\frac{(1-\nu)a}{R_0\phi_0 d}}$ (with $a=2R_0\alpha_0L$ the particle area)
and thus for fixed $\eta_0$ and $\nu$ we can compute the period $d$ from the latter equation. The integrals of Eq. (\ref{ooz}) 
were evaluated using a Gauss-Hermite quadrature while the minimization was carried out using the Newton-Raphson method 
with supplied numerical derivatives.  
 
\section{Results}
\label{secB}

\begin{figure}
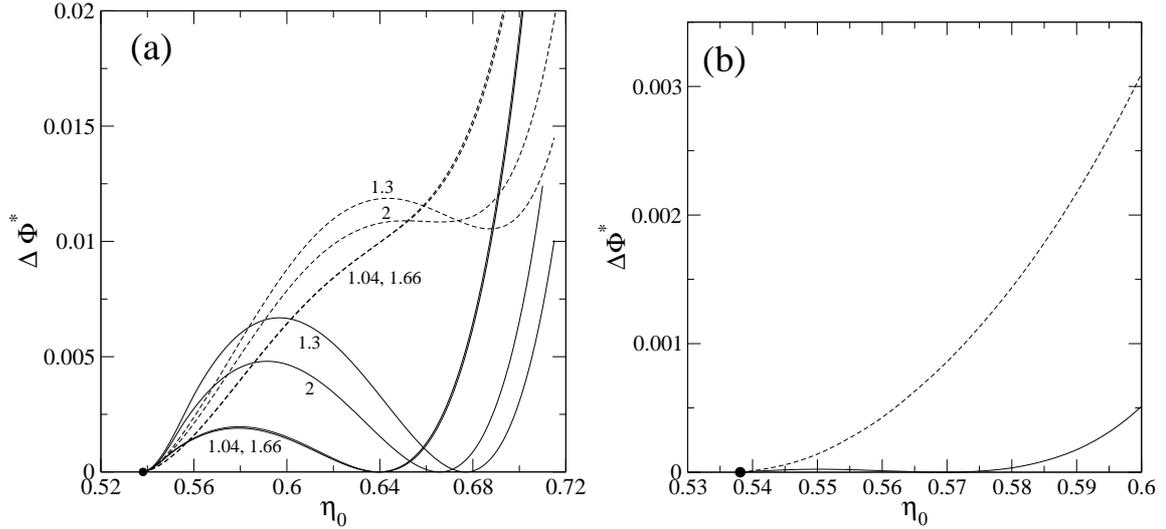

\epsfig{file=fig4a.eps,width=3.in}
\epsfig{file=fig4b.eps,width=2.9in}
\caption{(a): Free-energy differences 
$\Delta\Phi^*_{\rm C,K}$ 
(see the text for the definition) between the C and S (solid lines) phases 
and K and S phases (dashed lines) and for different values of $R_0/R$, in (a) they are 
equal to 1.04, 1.3, 1.66 and 2. Different lines are labelled with their corresponding values of $R_0/R$. 
The number of columns $N_0$ in the C phases are 2, 3, 4 and 5, respectively. In (b) $R_0/R=1.05$ and $N_0=2$. 
The crystalline phase in (b) was calculated through the free-energy
minimization with respect to the Fourier amplitudes of the density-profile Fourier expansion.}
\label{fig2}
\end{figure}

We minimized (\ref{todito}) using the Fourier expansion of the density profile (\ref{fourier}) for the S and C 
phases and for four different values of $R_0/R=1.04$, 1.30, 1.66 and 2.00. The results are shown in Fig. \ref{fig2} (a) in which we plot the difference 
between the free-energies per unit area corresponding to the C and S phases i.e. $\Delta\Phi^*_{\rm C}=\Phi^*_{\rm C}-\Phi^*_{\rm S}$ (with 
$\displaystyle{\Phi^*=\frac{\beta {\cal F} [\rho_{\rm eq}] a}{2\pi R_0{\cal L}}}$), as a function of the mean packing fraction $\eta_0$. 
For the C phase the number of columns for each value of $R_0/R$ 
are $N_0=2$, 3, 4 and 5, respectively. As can be seen, 
$\Delta\Phi^*_{\rm C}\geq 0$ always, implying that the S phase is the most stable phase. This behavior can be understood 
if we take into account that 
the C period in reduced units is $\phi_0^*=\phi_0 R_0/(2\alpha_0 R_0)=\pi/(N_0\alpha_0)$ which in turn is different 
from that of PHR (remember that the excess part of the free-energy density of CR can be obtained from that 
of PHR by the mapping $\sigma\to 2\alpha_0 R_0$, with $\sigma$ the width of the rectangle); the latter is calculated 
by minimizing the free-energy with respect to the period.  Note that, for PHR, the C and S phases have the same free energies (the 
PHR can be obtained by scaling the PHS along one direction). 
However in the present case the value of $\phi_0^*$ 
is imposed once we fix $R_0/R$ and the number of columns $N_0$, the latter being dictated by the commensuration of a CA of length 
$2\alpha_0 R_0$ in a circle of perimeter $2\pi R_0$, i.e. $2\leq N_0\leq [\pi/\alpha_0]$. 
In a case when different values of $N_0$ are possible, we select the one
which minimizes the free energy. It is interesting to note that $\Delta\Phi^*=0$ both at the bifurcation point $\eta_0^*$ and at that value 
of $\eta_0$ for which the periods in reduced units of CR ($\phi_0^*$) and of hard parallel rectangles (HPR) on a plane
($d^*_{\rm HPR}\equiv d_{\rm HPR}/\sigma$) are exactly the same. Finally, from Fig. \ref{fig2}(a), we can see that 
$\Delta\Phi^*_{\rm c}$ for the cases $R_0/R=1.04$ ($N_0=2$) and $1.66$ ($N_0=4$) are indistinguishable from each other.
The reason behind this behavior is that the commensuration numbers $\pi/(N_0\alpha_0)$ 
[ratio between the perimeter of 
the circle ($2\pi R_0$) and the total length of the $N_0$ arcs of length $2R_0\alpha_0$] are approximately equal in both cases 
(1.2153 and 1.2148, respectively).

\begin{figure}
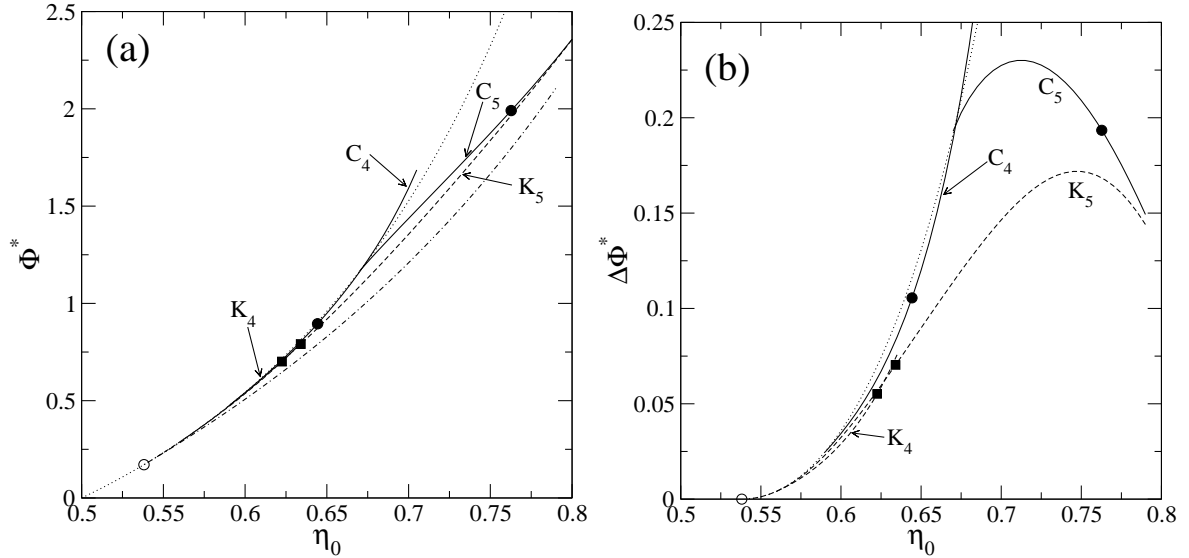

\epsfig{file=fig5a.eps,width=3.in}
\epsfig{file=fig5b.eps,width=3.in}
\caption{(a): Free energies per unit of area (we plot $\Phi^*+\eta_0\ln (a)$) of N (dotted), S (dot-dashed), 
C$_4$ and C$_5$ (solid lines correspondingly labelled), K$_4$ and 
K$_5$ (dashed lines  correspondingly labelled) phases as a function of $\eta_0$ for $R_0/R=1.8$. 
The open circle indicates the N-S bifurcation point. Circles indicate
the metastable C$_4$-C$_5$ coexistence, while squares correspond to the metastable K$_4$-K$_5$ coexistence. (b): 
Free-energy density differences between the S and all the phases shown in (a). The lines and symbols have the same 
meaning as in (a).} 
\label{fig4}
\end{figure}

In the same figure we also plot $\Delta\Phi^*_{\rm K}=\Phi^*_{\rm K}-\Phi^*_{\rm S}$, 
i.e. the difference between the free energies per unit volume of the K and S phases. 
For this case we have used the Gaussian parametrization (\ref{gaussian}) and minimized (\ref{ooz}) with respect to
$\Lambda$ and $d^*$ [we tried to use the Fourier expansion (\ref{fourier}) with K symmetry,  
but the numerical algorithm converged to a solution with C or S symmetry always, except for those values of $\eta_0$ 
close enough to the bifurcation point, an example of which  
is shown in Fig. \ref{fig2} (b)]. 
In Fig. \ref{fig2} (a) we see that the K phase has a larger free energy compared to the S phase. However now $\Delta\Phi^*_{\rm K}$ does not 
touch the $\eta_0$-axis tangentially, because 
the PHR free energies corresponding to the C (or S) and K phases are not equal, 
the former being energetically favored up to $\eta_0 \sim 0.73$ \cite{Rene}. 

It is interesting to note that the N-S transition exists even for values of $R_0/R\gtrsim 1$. As already shown, for 
$R_0/R< 1$ this transition is absent since the free energy functional is in fact the 1D functional for hard rods
(Percus functional).     

Now we present the results for $R_0/R=1.8$, a value for which the free-energy branches of the C phases 
with $N_0=4$ (C$_4$) and and $5$ (C$_5$) columns intersect at some value of $\eta_0$. 
These free energies are plotted in Fig. \ref{fig4}. The C$_4$ and C$_5$ branches
are above those of the K$_{4,5}$ and S phases. This result could change in a confined binary fluid of HC with different
lengths. For dissimilar enough lengths, the S and K phases would be energetically unfavored with respect 
to the C$_{N_0}$ phases. In this situation a $C_4$-$C_5$ transition could take place as $\eta_0$ increases. For some higher values 
of $R_0/R$ we will find a cascade of $C_{N_0-1}\to C_{N_0}$ transitions for different values of $N_0$ because  
multiple values of $N_0$ fulfills the constraint $2\leq N_0\leq [\pi/\alpha_0]$.     

\begin{figure}
\epsfig{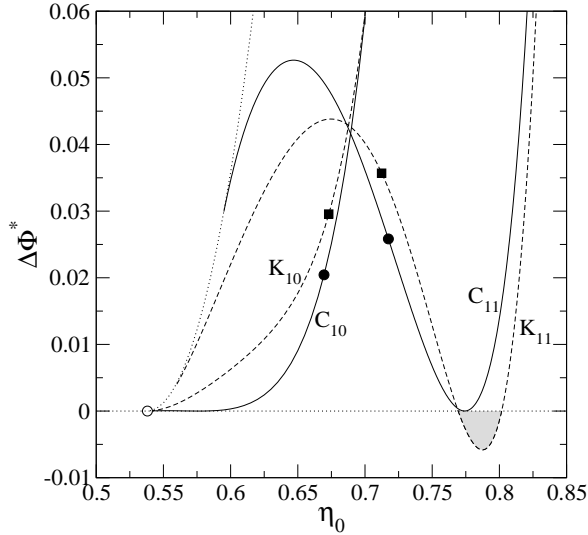}
\caption{Free-energy density differences between the S and the N (dotted), C$_{10}$ and
C$_{11}$ (solid lines correspondingly labelled), K$_{10}$ and
K$_{11}$ (dashed lines correspondingly labelled) phases as a function of the mean packing fraction $\eta_0$ for $R_0/R=4$.
The open circle indicates the N-S bifurcation point.
The circles and squares show the C$_{10}$-C$_{11}$ and K$_{10}$-K$_{11}$ metastable coexistences, respectively. 
The grey region shows the packing
fraction interval in which the K$_{11}$ has a lower free energy than the S phase.}
\label{the_last}
\end{figure}

Finally, we have calculated all the energy branches of stable or metastable phases 
for $R_0/R=4$, which are plotted in Fig. \ref{the_last}. For this case there are two columnar, C$_{10}$ and C$_{11}$, and 
two crystalline, K$_{10}$ and K$_{11}$, phases which are metastable with respect to the Sm phase 
(except the K$_{11}$ phase, which is stable in a small range of $\eta_0$). We show in Fig. \ref{the_last} 
the metastable C$_{10}$-C$_{11}$ and K$_{10}$-K$_{11}$ coexistences, using
different symbols. Note that the coexistence gaps decrease with respect to
the case $R_0/R=1.8$, a trend that should be confirmed if the value $R_0/R$ were to
be increased even more. 
It is interesting to note that there is a 
relatively small interval of packing fractions (dashed region in the figure) where the K$_{11}$ phase becomes reentrant.
This interval should increase 
for higher values of $R_0/R$ and n fact, in the limit $R_0/R\to\infty$ (the HPR limit), the K phase becomes stable for $\eta\gtrsim 0.73$ \cite{Rene}. 

\section{Density functional of spherical lenses}
\label{lenses}

In principle the same dimensional cross-over procedure can be implemented on the density functional for HS of 
radii $R$ whose centers of mass are restricted to be
on a spherical surface of radius $R_0$. If the density functional for HS adequately fulfills the dimensional cross-over property, the resulting
functional, that of a fluid consisting of hard spherical lenses (SL) [obtained from the intersection between the spherical surface and
a cone with vertex at the origin and with solid angle equal to $4\pi\sin^2(\alpha_0/2)=2\pi\left[1-\sqrt{1-(R/R_0)^2}\right]$],
should coincide, when $R_0/R>1$, with the 2D functional for HD [Eq. (\ref{oo})] where the density
profile this time is a periodic function of the spherical angles
$\rho(\phi,\theta)=\rho(\phi+2\pi,\theta+2\pi)$ and the semi-length of the SL is
equal to $R_0\alpha_0=R_0\sin^{-1}(R/R_0)$.
We present in Appendix \ref{a_e} the explicit expressions for the weighted densities of SL. With the present tool, the study of the freezing of HS on a spherical surface could be carried out. An interesting point to be studied is how the symmetry
of the crystalline structure, and the presence of defects and vacancies, could change with the ratio $R_0/R$.

\section{Conclusions}
\label{SecIII}
 The present work followed two motivations. The first was to illustrate the
dimensional cross-over criterion as a powerful projection tool to obtain a DF
for hard particles with centres of mass constrained to be on a particular (in
the present case cylindrical) surface from a given DF which fulfills this
criterion. The second motivation was the study of the phase behaviour of CR
(the particles obtained by projecting the centres of mass of HC onto
a cylindrical surface). We show that surface curvature has a profound impact on
the stability of the liquid-crystal phases of curved rectangles as compared with
the planar case. The projection method is quite general and can be used to
obtain DFs for other particle geometries and other surfaces. As an example, we
proposed in Sec. \ref{lenses} a DF for spherical lenses, i.e. the particles obtained
when the centres of mass of HS are placed on an external spherical surface.

During the derivation of DF for CR 
we proved the dimensional cross-over properties of the FMF of HD of radii $R$ 
when their centers of mass are constrained 
to be on a circumference of radius $R_0$. Depending on the ratio $R_0/R$ the original 2D density functional reduces to the 1D Percus 
functional (when $R_0/R>1$) or to the 0D functional ($R_0/R<1$), which are both exact limits. 
From these dimensionally-reduced functionals, using the dimensional expansion procedure, we derive the density functional 
for CR moving on a
cylindrical surface of radius $R_0$. We show that this functional is equivalent to that
of PHR on a flat surface with the edge-lengths of the particles being $\sigma=2R_0\sin^{-1}(R/R_0)$ and $L$.
We minimized the functional for CR to get 
the phase behavior for different values of $R_0/R$. When $R_0/R>1$ we obtain that the most stable phase is S, as
compared to the C or K phases, 
except for some relatively large values of $R_0/R$ and small density intervals in which we find a reentrant K phase. 
We find also metastable 
C$_{n-1}$-C$_n$ or K$_{n-1}$-K$_{n}$ transitions related to the commensuration between the particle width and the perimeter 
of the cylindrical surface. These transitions could become stable when the confined HC have different lengths (it is well known that 
the length 
polydispersity destroy the S ordering). When $R_0/R<1$ the density functional of CR coincides with the 1D Percus functional, so that
the system does not exhibit any phase transition. We are presently performing MC simulations of CR with the aim to compare
the phase behavior with 
that obtained from the present FMF. Note that the latter, being a 2D functional, is not exact. 

The present phase behavior points to possible textures 
that adsorbed molecules (for example proteins) on cylindrical membranes   
could exhibit. If these molecules are highly anisotropic, highly oriented along the cylinder axis and  
interact repulsively with each other, their stable textures should include only nematic and smectic-like configurations for low 
and high densities, respectively. 

The wall of some rod-shaped bacteria grows in the direction of the cylinder axis, keeping the radius 
approximately constant. The new proteins come from the inside of the cell to the wall. In Ref. \cite{Nelson3} the authors 
explain cell growth by the insertion of these proteins into the wall and their subsequent active diffusion 
along the perimeter of the cylinder via 
a dislocation-mediated growth. The model assumes that the proteins form a 
square lattice on a surface of the cylinder (this is apparently confirmed by experiments). 
Protein diffusion along the wall-perimeter 
(activated by the cell machinery) 
departs from some of these dislocations which constitute the source of the new proteins coming from the cell.
Our results show that the smectic-like configuration of proteins favors their transversal diffusion. Also the sources of 
new molecules could be located at any position inside the smectic layers with marginal energy cost associated 
to the spatial deformation of layers (as they are fluid-like). However this mechanism makes the cylindrical membrane grow
in the transversal direction. Growth along the longitudinal direction is possible by creating new smectic layers and 
consequently smectic-like defects in the layers. 
This is an alternative mechanism that might explain the growth of cells whose membranes are constituted by molecules that form smectic-like textures.

\appendix

\section{$2D\to 1D,0D$ dimensional cross-over of HD on a circle}
\label{a_a}
Substitution of (\ref{rho}) into Eq. (\ref{n0}) gives 
\begin{eqnarray}
n_0^{(\rm HD)}({\bf r})=\frac{R_0}{2\pi R}\int_0^{2\pi}d\phi_1 \delta(R-|{\bf r}-R_0 {\bf u}_1|)\rho_{\rm 1D}(\phi_1),
\label{n00}
\end{eqnarray}
where ${\bf u}_1=(\cos\phi_1,\sin\phi_1)$ is a unit vector. Here the polar radius $r$ is restricted to be in
the interval $[|R_0-R|,R_0+R]$. Using the change of variables
$\xi=|{\bf r}-R_0{\bf u}_1|=\sqrt{r^2+R_0^2-2rR_0\cos(\phi-\phi_1)}$, where 
${\bf r}=r{\bf u}$, ${\bf u}=(\cos\phi,\sin\phi)$ and ${\bf u}\cdot {\bf u}_1=\cos(\phi-\phi_1)$, 
we obtain 
\begin{eqnarray}
n_0^{(\rm HD)}({\bf r})=\frac{1}{2\pi r}\frac{\left[\rho_{\rm 1D}(\phi+\alpha(r))+\rho_{\rm 1D}(\phi-\alpha(r))\right]}
{\displaystyle{\sqrt{1-\left(\frac{r^2+R_0^2-R^2}{2rR_0}\right)^2}}}\Theta (R-|r-R_0|),
\label{n02}
\end{eqnarray}
where 
\begin{eqnarray}
\alpha(r)\equiv\cos^{-1}\left(\frac{r^2+R_0^2-R^2}{2rR_0}\right)
\end{eqnarray}
Now we calculate the two-body weighted density $N_{\rm HD}({\bf r})$ by substituting $\rho_{\rm 2D}({\bf r}_i)=\rho_{\rm 1d}(\phi_i)
\delta(R_0-r_i)$ [which is equivalent to setting ${\bf r}_i=R_0 {\bf u}_i$ with ${\bf u}_i=(\cos\phi_i,\sin\phi_i)$] into Eq. (\ref{two-body}).
Noting that  $r_{12}=|{\bf r}_1-{\bf r}_2|=
2R_0|\sin(\phi_{12}/2)|$ and using Eq. (\ref{kernel}), we obtain
\begin{eqnarray}
{\cal K}(r_{12})=4\pi R^2\frac{R_0}{R}|\sin(\phi_{12}/2)|
\sin^{-1}\left(\frac{R_0}{R}|\sin(\phi_{12}/2)|\right)
\sqrt{1-\left(\frac{R_0}{R}\sin(\phi_{12}/2)\right)^2}.
\end{eqnarray}
Using again $\xi_i=|{\bf r}-{\bf r}_i|$, 
taking into account that $\displaystyle{|\sin(\phi_{12}/2)|_{\rm \xi_i=R}=|\sin\alpha(r)|}$, 
and using (\ref{two-body}), we obtain
\begin{eqnarray}
N_{\rm HD}({\bf r})&=&\frac{R_0}{\pi r^3} \frac{\rho_{\rm 1D}(\phi-\alpha(r))\rho_{\rm 1D}(\phi+\alpha(r))}
{\displaystyle{\sqrt{1-\left(\frac{r^2+R_0^2-R^2}{2rR_0}\right)^2}}}|R^2+r^2-R_0^2|\nonumber\\
&&\times 
\sin^{-1}\left(\frac{R_0}{R}\sqrt{1-\left(\frac{r^2+R_0^2-R^2}{2rR_0}\right)^2}\right)
\Theta(R-|r-R_0|).\label{two-body2}
\end{eqnarray}
The local packing fraction (\ref{local}) results in
\begin{eqnarray}
\eta_{\rm HD}({\bf r})=R_0\int_0^{2\pi}d\phi_1\rho_{\rm 1D}(\phi_1)\Theta(R-|{\bf r}-R_0{\bf u}_1|)=
R_0\int_{\phi-\alpha(r)}^{\phi+\alpha(r)}d\phi_1\rho_{\rm 1D}(\phi_1),\label{eta_ast}
\end{eqnarray}
where we used the fact that the condition $|{\bf r}-R_0{\bf u}_1|\leq R$ implies $|\phi-\phi_1|\leq \alpha(r)$.
The weighted densities can be written in terms of the angle $\alpha(r)$ and 
the new angles $\theta(r)$ and $\gamma(r)$, defined in Fig. \ref{fig1}:
\begin{eqnarray}
&&n_0^{(\rm HD)}({\bf r})=\frac{R_0}{2\pi r}\left[\rho_{\rm 1D}(\phi-\alpha(r))+\rho_{\rm 1D}(\phi+\alpha(r))\right]\frac{d\gamma}{dr}(r),
\label{new_n0}\\\nonumber\\
&&N_{\rm HD}({\bf r})= \frac{2R_0^2}{\pi r}\rho_{\rm 1D}(\phi-\alpha(r))\rho_{\rm 1d}(\phi+\alpha(r))
\left|\frac{d\alpha}{d\gamma}(r)\right|\theta^*(r)\frac{d\gamma}{dr}(r),
\label{new_N}
\end{eqnarray}
where 
\begin{eqnarray}
\theta^*=\left\{
\begin{array}{ccc}
\theta, & \text{if} & 0\leq\theta\leq\pi/2\\
\pi-\theta, & \text{if} & \pi/2\leq \theta\leq\pi.
\end{array}
\right.
\end{eqnarray}
The following equation is satisfied:
\begin{eqnarray}
\rho_{\rm 1D}(\phi-\alpha(r))\rho_{\rm 1D}(\phi+\alpha(r))=
\frac{1}{4}\left\{\frac{2}{R_0}n_{0+}({\bf r})\frac{\partial \eta_{\rm HD}}{\partial\alpha}({\bf r})
-\frac{1}{R_0^2}\left[\frac{\partial \eta_{\rm HD}}{\partial\phi}({\bf r})\right]^2\right\}.
\label{ss0}
\end{eqnarray}
This can be derived from the relation
\begin{eqnarray}
\frac{\partial \eta_{\rm HD}}{\partial \alpha}({\bf r})=2R_0n_{0+}({\bf r}),\quad 
\frac{\partial \eta_{\rm HD}}{\partial \phi}({\bf r})=2R_0
n_{0-}({\bf r}),
\end{eqnarray}
with
\begin{eqnarray}
n_{0\pm}({\bf r})=\frac{1}{2}\left[\rho_{\rm 1D}(\phi+\alpha(r))\pm\rho_{\rm 1D}(\phi-\alpha(r))\right].
\end{eqnarray}
Now using the relations
\begin{eqnarray}
&&\frac{2}{R_0}\frac{\partial n_{0+}}
{\partial\alpha}({\bf r})-\frac{1}{R_0^2}\frac{\partial^2\eta_{\rm HD}}{\partial\phi^2}({\bf r})=0,\nonumber\\\nonumber\\
&&\frac{\displaystyle{n_{0+}({\bf r})\frac{\partial \eta_{\rm HD}}{\partial\alpha}({\bf r})}}{1-\eta_{\rm HD}({\bf r})}=
-\frac{\partial }{\partial \alpha}
\left\{n_{0+}({\bf r})\ln[1-\eta_{\rm HD}({\bf r})]\right\}+\frac{\partial n_{0+}}{\partial\alpha}({\bf r})
\ln[1-\eta_{\rm HD}({\bf r})],\nonumber\\\nonumber\\
&&\frac{\displaystyle{\left[\frac{\partial \eta_{\rm HD}}{\partial\phi}({\bf r})\right]^2}}{1-\eta_{\rm HD}({\bf r})}=
-\frac{\partial}{\partial\phi}\left\{\frac{\partial \eta_{\rm HD}}{\partial\phi}({\bf r})
\ln[1-\eta_{\rm HD}({\bf r})]\right\}+
\frac{\partial^2\eta_{\rm HD}}{\partial \phi^2}({\bf r})\ln[1-\eta_{\rm HD}({\bf r})]
\end{eqnarray}
we get
\begin{eqnarray}
&&\frac{\displaystyle{2n_{0+}({\bf r})\frac{\partial \eta_{\rm HD}}{\partial\alpha}({\bf r})}}
{R_0[1-\eta_{\rm HD}({\bf r})]}-\frac{\displaystyle{\left[\frac{\partial \eta_{\rm HD}}{\partial\phi}({\bf r})\right]^2}}
{R_0^2\left[1-\eta_{\rm HD}({\bf r})\right]}\nonumber\\\nonumber\\
&&=-\frac{2}{R_0}\frac{\partial}{\partial\alpha}\left\{n_{0+}({\bf r})\ln[1-\eta_{\rm HD}({\bf r})]\right\}+
\frac{1}{R_0^2}\frac{\partial}{\partial \phi}\left\{\frac{\partial \eta_{\rm HD}}{\partial\phi}({\bf r})\ln[1-\eta_{\rm HD}({\bf r})]\right\}.
\label{ss2}
\end{eqnarray}
Next we take into account the periodic condition  
$\eta_{\rm HD}(r,\phi)=\eta_{\rm HD}(r,\phi+2\pi)$ which implies
\begin{eqnarray}
\int_0^{2\pi}d\phi\frac{\partial}{\partial\phi}\left\{\frac{\partial \eta_{\rm HD}}{\partial \phi}({\bf r})\ln[1-\eta_{\rm HD}({\bf r})]
\right\}=0.
\label{condition}
\end{eqnarray}
Eqs. (\ref{ss0}), (\ref{ss2}) and (\ref{condition}) can be used to give
\begin{eqnarray}
\int_0^{2\pi}d\phi \frac{2R_0^2}{\pi}\frac{\rho_{\rm 1D}(\phi-\alpha(r))\rho_{\rm 1D}(\phi+\alpha(r))}{1-\eta_{\rm HD}({\bf r})}
=-\int_0^{2\pi}d\phi\frac{R_0}{\pi}\frac{\partial }{\partial\alpha}\left\{n_{0+}({\bf r})\ln[1-\eta_{\rm HD}({\bf r})]\right\}.
\label{last}
\end{eqnarray}
Introducing the change of variable $(r,\phi)\to(\gamma,\phi)$ 
and using Eqs. (\ref{new_n0}), (\ref{new_N}) and (\ref{last}), the excess part of the free-energy 
(\ref{excess}) can be rewritten as (\ref{ener}).

\subsection{The case $R_0>R$}
\label{a_b}

Using (\ref{angulos}) we have
\begin{eqnarray}
&&\int_0^{\pi}d\gamma\frac{\partial\Psi}{\partial\alpha}(\phi,\alpha(\gamma))\theta^*(\gamma)\left|
\frac{d\alpha}{d\gamma}(\gamma)\right|=
\int_0^{\gamma_0}d\gamma\frac{\partial\Psi}{\partial \alpha}(\phi,\alpha(\gamma))(\pi-\theta(\gamma))
\frac{d\alpha}{d\gamma}(\gamma)\nonumber\\\nonumber\\
&&+\int_{\gamma_0}^{\pi}d\gamma
\frac{\partial\Psi}{\partial \alpha}(\phi,\alpha(\gamma))
\theta(\gamma) \left(-\frac{d\alpha}{d\gamma}(\gamma)\right)\nonumber\\\nonumber\\
&&=\left[\pi \Psi(\phi,\alpha)\right]_{\alpha(0)}^{\alpha_0}-\underbrace{\int_0^{\pi}d\gamma
\frac{\partial}{\partial\alpha}\left[\Psi(\phi,\alpha(\gamma))\theta(\gamma)\right]\frac{d\alpha}{d\gamma}(\gamma)}_{
=\left[\Psi(\phi,\alpha)\theta(\alpha)\right]_{\alpha(0)}^{\alpha(\pi)}=0}
+\int_0^{\pi}d\gamma \Psi(\phi,\alpha(\gamma))\frac{d\theta}{d\alpha}(\gamma)\frac{d\alpha}{d\gamma}(\gamma)\nonumber\\\nonumber\\
&&=\pi \Psi(\phi,\alpha_0)+\int_0^{\pi}d\gamma \Psi(\phi,\alpha(\gamma))\frac{d\theta}{d\alpha}(\gamma)
\frac{d\alpha}{d\gamma}(\gamma), \label{lala}
\end{eqnarray}
where we have used the change of variable $\displaystyle{\int_{\gamma_1}^{\gamma_2}d\gamma T(\phi,\alpha(\gamma))
\frac{d\alpha}{d\gamma}(\gamma)=\int_{\alpha(\gamma_1)}^{\alpha(\gamma_2)} d\alpha T(\phi,\alpha)}$ 
for a general function $T(\phi,\alpha)$ and also that $\alpha(0)=\alpha(\pi)=0$.
Now from (\ref{lala}) and (\ref{ener}) we can reexpress the excess free energy as
\begin{eqnarray}
\beta{\cal F}_{\rm ex}^{(\rm HD)}[\rho]=\frac{R_0}{\pi}\int_0^{2\pi}d\phi\left\{\pi \Psi(\phi,\alpha_0)
+\int_0^{\pi}d\gamma \Psi(\phi,\alpha(\gamma))\left(1+\frac{d\theta}{d\alpha}(\gamma)\frac{d\alpha}{d\gamma}(\gamma)\right)\right\}.
\end{eqnarray}
Furhter, since $\alpha+\theta+\gamma=\pi$, we have
\begin{eqnarray}
d\gamma+\frac{d\theta}{d\alpha}\frac{d\alpha}{d\gamma}d\gamma=d\alpha\left(\frac{d\gamma}{d\alpha}
+\frac{d\theta}{d\alpha}\right)=-d\alpha
\label{triangle}
\end{eqnarray}
and
\begin{eqnarray}
\beta{\cal F}_{\rm ex}^{(\rm HD)}[\rho]=\frac{R_0}{\pi}\int_0^{2\pi}d\phi\left\{\pi\Psi(\phi,\alpha_0) 
-\underbrace{\int_{\alpha(0)}^{\alpha(\pi)}d\alpha \Psi(\phi,\alpha)}_{=0}\right\}.
\end{eqnarray}
Let us redefine the weighted densities as in (\ref{redefine_1}) and (\ref{redefine_2})
where the index `CA' means that these densities are those on a circular arc of length $2R_0\alpha_0$. Then we have proved 
that the excess part of the HD free-energy functional reduces to that of hard CA, given by  Eq. (\ref{limite1}).

\subsection{The case $R_0<R$}
\label{a_c}

Taking into account (\ref{angles2}), Eq. (\ref{ener}) reads 
\begin{eqnarray}
\beta {\cal F}_{\rm ex}^{(\rm HD)}[\rho] =
\frac{R_0}{\pi}\int_0^{2\pi}d\phi\int_0^{\pi}d\gamma \left\{\Psi(\phi,\alpha(\gamma))-\frac{\partial\Psi}{\partial\alpha}
(\phi,\alpha(\gamma))\theta(\gamma) \frac{d\alpha}{d\gamma}(\gamma)\right\}.
\label{st}
\end{eqnarray}
Using
\begin{eqnarray}
\frac{\partial\Psi}{\partial\alpha}(\phi,\alpha)\theta(\gamma)=
\frac{\partial}{\partial\alpha}\left[\Psi(\phi,\alpha)\theta(\gamma)\right]-\Psi(\phi,\alpha)\frac{d\theta}
{d\alpha}(\gamma),
\end{eqnarray}
Eq. (\ref{st}) becomes 
\begin{eqnarray}
&&\beta{\cal F}_{\rm ex}^{(\rm HD)}[\rho]=
\frac{R_0}{\pi}\int_0^{2\pi}d\phi\left[\int_0^{\pi}d\gamma \Psi(\phi,\alpha(\gamma))
\left(1+\frac{d\theta}{d\alpha}(\gamma)
\frac{d\alpha}{d\gamma}(\gamma)\right)\right.\nonumber\\
&&\left.-\int_0^{\pi}d\gamma\frac{d\alpha}{d\gamma}(\gamma)\frac{\partial}{\partial\alpha}
\left(\Psi(\phi,\alpha(\gamma))\theta(\gamma)\right)\right],
\end{eqnarray}
and, using (\ref{triangle}),
\begin{eqnarray}
\hspace{-0.4cm}\beta{\cal F}_{\rm ex}^{(\rm HD)}[\rho]=\frac{R_0}{\pi}\int_0^{2\pi}d\phi
\left\{-\int_{\pi}^{0}d\alpha \Psi(\phi,\alpha)
-\underbrace{\left[\Psi(\phi,\alpha)\theta(\alpha)\right]_{\pi}^{0}}_{=0}\right\}=
\frac{R_0}{\pi}\int_0^{2\pi}d\phi\int_0^{\pi}d\alpha \Psi(\phi,\alpha),\nonumber\\
\label{gg}
\end{eqnarray}
where we used the fact that the function $\alpha(\gamma)$ has values
$\alpha(0)=\pi$ and $\alpha(\pi)=0$, while the function $\theta(\alpha)$ has values
$\theta(0)=\theta(\pi)=0$ (see Fig. \ref{fig1}).
Taking into account now that $\displaystyle{\frac{\partial \eta_{\rm HD}}{\partial\alpha}(\phi,\alpha)=
R_0\left[\rho_{\rm 1d}(\phi+\alpha)+\rho_{\rm 1d}(\phi-\alpha)\right]=2R_0n_{0+}(\phi,\alpha)}$, 
we obtain from (\ref{gg}):
\begin{eqnarray}
\beta{\cal F}_{\rm ex}^{(\rm HD)}[\rho]&=&
-\frac{R_0}{\pi}\int_0^{2\pi}d\phi\int_0^{\pi}d\alpha\frac{1}{2R_0}\frac{\partial \eta_{\rm HD}}{\partial\alpha}(\phi,\alpha)
\ln\left[1-\eta_{\rm HD}(\phi,\alpha)\right]\nonumber\\\nonumber\\
&=&\frac{1}{2\pi}\int_0^{2\pi}d\phi\int_0^{\pi}d\alpha\frac{\partial}{\partial\alpha}\left\{
\eta_{\rm HD}(\phi,\alpha)+[1-\eta_{\rm HD}(\phi,\alpha)]\ln[1-\eta_{\rm HD}(\phi,\alpha)]\right\}\nonumber\\\nonumber\\
&=&\frac{1}{2\pi}\int_0^{2\pi}d\phi\left[\eta_{\rm HD}(\phi,\alpha)+(1-\eta_{\rm HD}(\phi,\alpha))
\ln[1-\eta_{\rm HD}(\phi,\alpha)]\right]_0^{\pi}.
\label{finally}
\end{eqnarray}
Noting that $\eta_{\rm HD}(\phi,0)=0$ and that
\begin{eqnarray}
\eta_{\rm CA}\equiv 
\eta_{\rm HD}(\phi,\pi)=R_0\int_{\phi-\pi}^{\phi+\pi}d\phi'\rho_{\rm 1D}(\phi')=R_0\int_{-\pi}^{\pi}d\phi'
\rho_{\rm 1D}(\phi')=2\pi R_0 \overline{\rho}_{\rm 1D},
\end{eqnarray}
we arrive at (\ref{0DD}).

\section{Fourier and Gaussian parameterizations}
\label{a_d}

The expressions for the weighted densities of CR, using the Fourier expansion (\ref{fourier}), are 
\begin{eqnarray}
&&\eta_{\rm CR}(\phi,z)=\eta_0\left[1+\sum_{(k,m)\neq (0,0)}s_{km}
\chi_1(kN_0\alpha_0)\chi_1\left(\frac{m\pi}{d^*}\right)\cos(kN_0\phi)\cos(q m z)\right],\label{eta_f}\\\nonumber\\
&&n_0^{(\rm CR)}(\phi,z)=\rho_0\left[1+\sum_{(k,m)\neq (0,0)}s_{km}\chi_0(kN_0\alpha_0)\chi_0\left(\frac{m\pi}{d^*}\right)
\cos(kN_0\phi)\cos(q m z)\right],\label{n0_f}\\\nonumber\\
&&n_{1,\perp}^{(\rm CR)}(\phi,z)=\frac{\eta_0}{L}
\left[1+\sum_{(k,m)\neq (0,0)}s_{km}\chi_1(kN_0\alpha_0)\chi_0\left(\frac{m\pi}{d^*}\right)\cos(kN_0\phi)\cos(q m z)\right],\label{n1_per}\\\nonumber\\
&&n_{1,\parallel}^{(\rm CR)}(\phi,z)=\rho_0L\left[1+\sum_{(k,m)\neq (0,0)}s_{km}\chi_0(kN_0\alpha_0)\left(\frac{m\pi}{d^*}\right)
\cos(kN_0\phi)\cos(q m z)\right],\label{n1_par}
\end{eqnarray}
where $\eta_0=2\rho_0 R_0 \alpha_0 L$ is the mean packing fraction, $d^*=d/L$ is the $z$-period in reduced units, 
and $\chi_0(x)\equiv \cos x$, $\chi_1(x)\equiv\sin x/x$. 

Within the Gaussian parameterization (\ref{gaussian}), the free-energy density-functional of CR can be computed as
\begin{eqnarray}
\frac{\beta {\cal F}_{\rm CR}[\rho]a }{2\pi R_0 {\cal L}}&=&
\eta_0\left\{\ln\left[\frac{\eta_0\sqrt{\Lambda^*_{\perp}\Lambda^*_{\parallel}}}{\pi a}\right]-1
+\frac{2}{\sqrt{\pi}}\int_0^{\infty} dt e^{-t^2} 
\ln\left\{\prod_{\tau=\perp,\parallel} \left[\sum_k\exp\left(-\left(t-k\sqrt{\Lambda^*_{\tau}}\right)^2\right)\right]\right\}\right.
\nonumber\\\nonumber\\
&&+\left.\frac{1}{\pi}\int_0^{\infty}dt_1 e^{-t_1^2}\int_0^{\infty}dz e^{-t_2^2} T\left(\frac{t_1}{\sqrt{\Lambda_{\perp}}R_0},
\frac{t_2}{\sqrt{\Lambda_{\parallel}}}\right)\right\}.
\label{ooz}
\end{eqnarray}
where $a=2R_0\alpha_0 L$ is the particle area and 
we have defined $\Lambda^*_{\perp}\equiv\Lambda\left(\phi_0^*\right)^2$ [with $\phi_0^*=\pi/(N_0\alpha_0)$] , 
$\Lambda^*_{\parallel}\equiv\Lambda \left(d^*\right)^2$, and
\begin{eqnarray}
T(\phi,z)=\sum_{\tau_1,\tau_2=\pm} H(\phi_{\tau_1},z_{\tau_2}),\quad H(\phi,z)=-\ln \left[1-\eta_{\rm CR}(\phi,z)\right]
+\frac{\eta_{\rm CR}(\phi,z)}{1-\eta_{\rm CR}(\phi,z)},
\end{eqnarray}
with $\phi_{\pm}=\phi\pm\alpha_0$, $z=z\pm L/2$, and $\eta_{\rm CR}(\phi,z)=(1-\nu)\zeta_{1\perp}(\phi)\zeta_{1\parallel}(z)$. We 
have used the notation  
\begin{eqnarray}
&&\zeta_{1\perp}\left(\frac{t_1}{\sqrt{\Lambda_{\perp}}R_0}\right)=\frac{1}{2}\sum_k\left\{\text{erf}\left[t_1+\sqrt{\Lambda^*_{\perp}}
\left(\frac{1}{2\phi_0^*}-k\right)\right]-
\text{erf}\left[t_1-\sqrt{\Lambda^*_{\perp}}\left(\frac{1}{2\phi_0^*}+k\right)\right]\right\},\nonumber\\\nonumber\\
&&\zeta_{1\parallel}\left(\frac{t_2}{\sqrt{\Lambda_{\parallel}}}\right)=\frac{1}{2}\sum_k\left\{\text{erf}\left[t_2+\sqrt{\Lambda^*_{\parallel}}
\left(\frac{1}{2d^*}-k\right)\right]-
\text{erf}\left[t_2-\sqrt{\Lambda^*_{\parallel}}\left(\frac{1}{2d^*}+k\right)\right]\right\},
\end{eqnarray}
with $\text{erf}(x)$ the error function.

\section{DF for spherical lenses}
\label{a_e}

The correct dimensional cross-over of HS confined on a spherical surface means the following: When the 
density profile of HS is restricted to be on a spherical surface of radius $R_0$,
$\rho_{\rm 3D}(r,\hat{\boldsymbol{\Omega}})=\rho_{\rm 2D}(\hat{\boldsymbol{\Omega}})\delta(R_0-r)$, 
where $(r,\theta,\phi)$  
are the radius and the angles of spherical coordinates, while $\hat{\boldsymbol{\Omega}}(\theta,\phi)=
\left(\sin\theta\cos\phi,\sin\theta\sin\phi,\cos\theta\right)$ is the unit vector in the radial 
direction, we should obtain   
\begin{eqnarray}
\beta{\cal F}_{\rm ex}^{(\rm HS)}[\rho]\to \beta{\cal F}_{\rm ex}^{(\rm SL)}[\rho]=R_0^2\int 
d\hat{\boldsymbol{\Omega}}\Phi_{\rm 2D}^{(\rm SL)}(\hat{\boldsymbol{\Omega}}),
\end{eqnarray}
which results in a dimensionally-reduced density functional for spherical lenses (SL), where $d\hat{\boldsymbol{\Omega}}=
d\phi d\theta\sin\theta$ is the solid-angle element.
The expression for $\Phi_{\rm 2D}(\hat{\boldsymbol{\Omega}})$ is the same as in (\ref{oo}), with the weighted densities now defined as
\begin{eqnarray}
&& n_0^{(\rm SL)}(\hat{\boldsymbol{\Omega}})=\frac{R_0}{2\pi R}\int d\hat{\boldsymbol{\Omega}}_1\rho_{\rm 2D}(\hat{\boldsymbol{\Omega}}_1)
\delta\left(\alpha_0-\gamma(\hat{\boldsymbol{\Omega}},\hat{\boldsymbol{\Omega}}_1)\right),\\
&&\eta_{\rm SL}(\hat{\boldsymbol{\Omega}})=R_0^2\int d\hat{\boldsymbol{\Omega}}_1\rho_{\rm 2D}(\hat{\boldsymbol{\Omega}}_1)
\Theta\left(\alpha_0-\gamma(\hat{\boldsymbol{\Omega}},\hat{\boldsymbol{\Omega}}_1)\right),\\
&&N_{\rm SL}(\hat{\boldsymbol{\Omega}})=\left(\frac{R_0}{2\pi R}\right)^2\int d\hat{\boldsymbol{\Omega}}_1\int 
d\hat{\boldsymbol{\Omega}}_2
\rho_{\rm 2D}(\hat{\boldsymbol{\Omega}}_1)\rho_{\rm 2D}(\hat{\boldsymbol{\Omega}}_2)
\delta\left(\alpha_0-\gamma(\hat{\boldsymbol{\Omega}},\hat{\boldsymbol{\Omega}}_1)\right)
\delta\left(\alpha_0-\gamma(\hat{\boldsymbol{\Omega}},\hat{\boldsymbol{\Omega}}_2)\right)\nonumber\\ 
&&\times {\cal K}\left(\frac{R\gamma(\hat{\boldsymbol{\Omega}_1},\hat{\boldsymbol{\Omega}}_2)}{\alpha_0}\right),
\end{eqnarray}
where $\alpha_0=\sin^{-1}(R/R_0)$ and we have used the notations
\begin{eqnarray}
&&\cos \gamma (\hat{\boldsymbol{\Omega}},\hat{\boldsymbol{\Omega}}_i)=
\sin\theta\sin\theta_i\cos(\phi-\phi_i)+\cos\theta\cos\theta_i,\quad i=1,2,\\
&&\cos \gamma(\hat{\boldsymbol{\Omega}}_1,\hat{\boldsymbol{\Omega}}_2)=
\sin\theta_1\sin\theta_2\cos(\phi_1-\phi_2)+\cos\theta_1\cos\theta_2,
\end{eqnarray}
which define the angles between the unit vectors $\hat{\boldsymbol{\Omega}}$ and 
$\hat{\boldsymbol{\Omega}}_i$ and $\hat{\boldsymbol{\Omega}}_1$ and $\hat{\boldsymbol{\Omega}}_2$ respectively. 
The kernel ${\cal K}(r)$ is that defined in Eq. (\ref{kernel}). The condition $\rho_{\rm 2D}(\theta+2\pi,\phi+2\pi)$ 
should be imposed, as the density profile is now a periodic function of $\phi$ and $\theta$.

\begin{acknowledgments}
We acknowledge financial support from Comunidad Aut\'onoma de Madrid (Spain) under
the R$\&$D Programme of Activities MODELICO-CM/S2009ESP-1691, and from MINECO (Spain)
under grants {MOSAICO}, FIS2010-22047-C01 and FIS2010-22047-C04.
\end{acknowledgments}

\end{document}